\documentclass[entropy,article,accept,moreauthors,dvipdfm,12pt,a4paper]{mdpi}

\usepackage{amsmath,amssymb,graphicx,amsthm}
\input epsf
\usepackage{float}
\usepackage{hyphenat}
\usepackage[sort&compress]{natbib}

\newcommand{\D}{{\mathrm{d}}}

 \lastpage{1019} \doinum{10.3390/e13050966}
\pubvolume{13} \pubyear{2011} \history{Received: 25 January 2011; in
revised form: 28 March 2011 / Accepted: 12 May 2011 / \linebreak
Published: 20 May 2011 / Corrected postprint 12 April 2014}

\Title{The Michaelis-Menten-Stueckelberg Theorem}

\Author{Alexander N. Gorban $^{1,\star}$ and Muhammad Shahzad $^{1,2}$}

\address{$^{1}$Department of Mathematics, University of Leicester, Leicester, LE1
7RH, UK; \\$^{2}$Department of Mathematics, Hazara University, Mansehra, 21300, Pakistan}
\corres{E-Mail: ag153@le.ac.uk.}

\abstract{We study chemical reactions with complex mechanisms under
two assumptions: (i) intermediates are present in small amounts
(this is the quasi-steady-state hypothesis or QSS) and (ii) they are
in equilibrium relations with substrates (this is the
quasiequilibrium hypothesis or QE). Under these assumptions, we
prove the generalized mass action law together with the basic
relations between kinetic factors, which are sufficient for the
positivity of the entropy production but hold even without
microreversibility, when the detailed balance is not applicable.
Even though QE and QSS produce useful approximations by themselves,
only the combination of these assumptions can render the possibility
beyond the ``rarefied gas'' limit or the ``molecular chaos''
hypotheses. We do not use any a priori form of the kinetic law for
the chemical reactions and describe their equilibria by
thermodynamic relations. The transformations of the intermediate
compounds can be described by the Markov kinetics because of their
low density ({\em low density of elementary events}). This
combination of assumptions was introduced by Michaelis and Menten in
1913.  In 1952, Stueckelberg used the same assumptions for the gas
kinetics and produced the remarkable semi-detailed balance relations
between collision rates in the Boltzmann equation that are weaker
than the detailed balance conditions but are still sufficient for
the Boltzmann $H$-theorem to be valid. Our results are obtained
within the Michaelis-Menten-Stueckelbeg conceptual framework. }

\begin{document}

\section{Introduction}

\subsection{Main Asymptotic Ideas in Chemical Kinetics}

There are several essentially different approaches to asymptotic and
scale separation in kinetics, and each of them has its own area of
applicability.

In chemical kinetics various fundamental ideas about asymptotical
analysis were developed \cite{Klonowski1983}: Quasieqiulibrium
asymptotic (QE), quasi steady-state asymptotic (QSS), lumping, and
the idea of limiting step.

Most of the works on nonequilibrium thermodynamics deal with the QE
approximations and corrections to them, or with applications of
these approximations (with or without corrections). There are two
basic formulation of the QE approximation: The thermodynamic
approach, based on entropy maximum, or the kinetic formulation,
based on selection of fast reversible reactions. The very first use
of the entropy maximization dates back to the classical work of
Gibbs \cite{Gibb}, but it was first claimed for a principle of
informational statistical thermodynamics by Jaynes \cite{Jaynes1}. A
very general discussion of the maximum entropy principle with
applications to dissipative kinetics is given in the review
\cite{Bal}. Corrections of QE approximation with applications to
physical and chemical kinetics were
developed~\cite{GKIOeNONNEWT2001,GorKar}.

QSS was proposed by Bodenstein in 1913 \cite{Bodenstein1913},
and the important Michaelis and Menten
work~\cite{MichaelisMenten1913} was published simultaneously.
It appears that no kinetic theory of catalysis is possible
without QSS. This method was elaborated into an important tool
for the analysis of chemical reaction mechanism and
kinetics~\cite{Semenov1939,Christiansen1953,Helfferich1989}.
The classical QSS is based on the {\em relative smallness of
concentrations} of some of the ``active'' reagents (radicals,
substrate-enzyme complexes or active components on the catalyst
surface)~\cite{BriggsHaldane1925,Aris1965,Segel89}.

Lumping analysis aims to combine reagents into
``quasicomponents'' for dimension \linebreak reduction
\cite{LumpWei1,LumpWei2,LumpLiRab1,LumpLiRab2}. Wei and Prater
\cite{Wei62} demonstrated that for (pseudo)monomolecular
systems there exist linear combinations of concentrations which
evolve in time independently. These linear combinations
(quasicomponents) correspond to the left eigenvectors of the
kinetic matrix: If $l K= \lambda l$ then \linebreak $\D
(l,c)/\D t= (l,c) \lambda $, where the standard inner product
$(l,c)$ is the concentration of a quasicomponent. They also
demonstrated how to find these quasicomponents in a properly
organized experiment.

This observation gave rise to a question: How to lump
components into proper quasicomponents to guarantee the
autonomous dynamics of the quasicomponents with appropriate
accuracy? Wei and Kuo studied conditions for exact
\cite{LumpWei1} and approximate \cite{LumpWei2} lumping in
monomolecular and pseudomonomolecular systems. They
demonstrated that under certain conditions a large
monomolecular system could be well-modelled by a lower-order
system.

More recently, sensitivity analysis and Lie group approach were
applied to lumping analysis \cite{LumpLiRab1,LumpLiRab2}, and
more general nonlinear forms of lumped concentrations were used
(for example, concentration of quasicomponents could be a
rational function of $c$).

Lumping analysis was placed in the linear systems theory and
the relationships between lumpability and the concepts  of
observability, controllability and minimal realization were
demonstrated \cite{LumpingOBservability}. The lumping
procedures were considered also as efficient techniques leading
to nonstiff systems and demonstrated the efficiency of the
developed algorithm on kinetic models of atmospheric chemistry
\cite{NonstiffAtmospheric2002}.  An optimal lumping problem can
be formulated in the framework of a mixed integer nonlinear
programming (MINLP) and can be efficiently solved with a
stochastic optimization method \cite{OptimalLumping2008}.

The concept of limiting step gives the limit simplification:
The whole network behaves as a single step. This is the most
popular approach for model simplification in chemical kinetics
and in many areas beyond kinetics. In the form of a {\it
bottleneck} approach this approximation is very popular from
traffic management to computer programming and communication
networks. Recently, the concept of the limiting step has been
extended to the asymptotology of multiscale reaction networks
\cite{GorbaRadul2008,GorRadZin2010}.

In this paper, we focus on the combination of the QE approximation
with the QSS approach.

\subsection{The Structure of the Paper}

Almost thirty years ago one of us published a book \cite{G1} with
Chapter 3 entitled ``Quasiequilibrium and Entropy Maximum''. A
research program was formulated there, and now we are in the
position to analyze the achievements of these three decades and
formulate the main results, both theoretical and applied, and the
unsolved problems. In this paper, we start this work and combine a
presentation of theory and application of the QE approximation in
physical and chemical kinetics with exposition of some new~results.

We start from the formal description of the general idea of QE and
its possible extensions. In Section~\ref{General}, we briefly
introduce main notations and some general formulas for exclusion of
fast variables by the QE approximation.

In Section~\ref{MMConfusion}, we present the history of the QE
and the classical confusion between the QE and the quasi steady
state (QSS) approximation. Another surprising confusion is that
the famous Michaelis-Menten kinetics was not proposed by
Michaelis and Menten in 1913 \cite{MichaelisMenten1913} but by
Briggs and Haldane \cite{BriggsHaldane1925} in 1925. It is more
important that Michaelis and Menten proposed another
approximation that is very useful in general theoretical
constructions. We described this approximation for general
kinetic systems. Roughly speaking, this approximation states
that any reaction goes through transformation of fast
intermediate complexes (compounds), which  (i) are in
equilibrium with the input reagents and (ii) exist in a very
small amount.

One of the most important benefits from this approach is the
exclusion of nonlinear kinetic laws and reaction rate constants
for nonlinear reactions. The nonlinear reactions transform into
the reactions of the compounds production. They are in a fast
equilibrium and the equilibrium is ruled by thermodynamics. For
example, when Michaelis and Menten discuss the production of
the enzyme-substrate complex ES from enzyme E and substrate S,
they do not discuss reaction rates. These rates may be unknown.
They just assume that the reaction $E+S \rightleftharpoons ES$
is in equilibrium. Briggs and Haldane involved this reaction
into the kinetic model. Their approach is more general than the
Michaelis--Menten approximation but for the Briggs and Haldane
model we need more information, not only the equilibrium of the
reaction $E+S\rightleftharpoons ES$ but also its  rates and
constants.

When compounds undergo transformations in a linear first order
kinetics, there is no need to include interactions between them
because they are present in very small amounts in the same
volume, and their concentrations are also small. (By the way,
this argument is not applicable to the heterogeneous catalytic
reactions. Although the intermediates are in both small amounts
and in a small volume, {\em i.e.}, in the surface layer, the
concentration of the intermediates is not small, and their
interaction does not vanish when their amount decreases
\cite{Yablonskii1991}. Therefore, kinetics of intermediates in
heterogeneous catalysis may be nonlinear and demonstrate
bifurcations, oscillations and other complex behavior.)

In 1952, Stueckelberg \cite{Stueckelberg1952} used similar approach
in his seminal paper ``$H$-theorem and unitarity of the
$S$-matrix''. He studied elastic collisions of particles as the
quasi-chemical reactions $$\mathbf{v+w \to v'+w'}$$
($\mathbf{v,w,v',w'}$ are velocities of particles) and demonstrated
that for the Boltzmann equation the linear Markov kinetics of the
intermediate compounds results in the special relations for the
kinetic coefficients. These relations are sufficient for the
$H$-theorem, which was originally proved by Boltzmann under the
stronger assumption of reversibility of collisions \cite{Boltzmann}.

First, the idea of such relations was proposed by Boltzmann as
an answer to the Lorentz objections against Boltzmann's proof
of the $H$-theorem. Lorentz stated the nonexistence of inverse
collisions for polyatomic molecules. Boltzmann did not object
to this argument but proposed the ``cyclic balance'' condition,
which means balancing in cycles of transitions between states
$S_1 \to S_2 \to \ldots \to S_n \to S_1$. Almost 100 years
later, Cercignani and Lampis \cite{CercignaniLamp1981}
demonstrated that the Lorenz arguments are wrong and the new
Boltzmann relations are not needed for the polyatomic molecules
under the microreversibility conditions. The detailed balance
conditions should hold.

Nevertheless, Boltzmann's idea is very seminal. It was studied
further by Heitler \cite{Heitler1944} and Coester~\cite{Coester1951}
and the results are sometimes cited as the ``Heitler-Coestler
theorem of semi-detailed balance''. In 1952,
Stueckelberg~\cite{Stueckelberg1952} proved these conditions for the
Boltzmann equation. For the micro-description he used the $S$-matrix
representation, which is in this case equivalent for the Markov
microkinetics (see also~\cite{Watanabe1955}).

Later, these relations for the chemical mass action kinetics were
rediscovered and called the {\em complex balance conditions}
\cite{HornJackson1972,Feinberg1972}. We generalize the
Michaelis-Menten-Stueckelberg approach and study in
Section~\ref{GeneralKinetics} the general kinetics with fast
intermediates present in small amount. In
Subsection~\ref{Sec:TheTeorem} the big
Michaelis-Menten-Stueckelberg theorem is formulated as the overall
result of the previous analysis.

Before this general theory, we introduce the formalism of the QE
approximation with all the necessary notations and examples for
chemical kinetics in Section~\ref{ChemKinQEApprox}.

The result of the general kinetics of systems with intermediate
compounds can be used wider than this specific model of an
elementary reaction: The intermediate complexes with fast
equilibria and the Markov kinetics can be considered as the
``construction staging'' for general kinetics. In
Section~\ref{Sec:GenKinThermod}, we delete the construction
staging and start from the general forms of the obtained
kinetic equations as from the basic laws. We study the
relations between the general kinetic law and the thermodynamic
condition of the positivity of the entropy production.

Sometimes the kinetics equations may not respect thermodynamics
from the beginning. To repair this discrepancy, deformation of
the entropy may help. In Section~\ref{Sec:EntropyDeform}, we
show when is it possible to deform the entropy by adding a
linear function to provide agreement between given kinetic
equations and the deformed thermodynamics. As a particular
case, we got the ``deficiency zero~theorem''.

The classical formulation of the principle of detailed balance
deals not with the thermodynamic and global forms we use but
just with equilibria: In equilibrium each process must be
equilibrated with its reverse process. In
Section~\ref{Sec:EntropyDeform}, we demonstrate also that for
the general kinetic law the existence of a point of detailed
balance is equivalent to the existence of such a linear
deformation of the entropy that the global detailed balance
conditions (Equation (\ref{detailed balance}) below) hold.
Analogously, the existence of a point of complex balance is
equivalent to the global condition of complex balance after
some linear deformation of the entropy.

\subsection{Main Results: One Asymptotic and Two Theorems}

\begin{figure}[H]
\centering{
\includegraphics[width=0.5\textwidth]{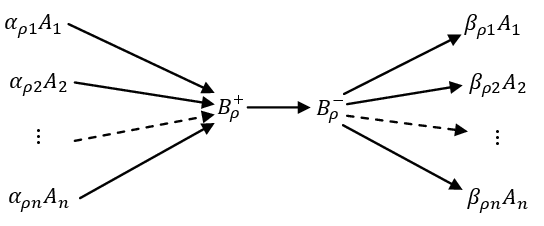}
}\caption{A $2n$-tail scheme of an extended elementary reaction.
\label{n-tail}}
\end{figure}

Let us follow the ideas of Michaelis-Menten and Stueckelberg and
introduce the asymptotic theory of reaction rates. Let the list of
the components $A_i$ be given. The mechanism of reaction is the list
of the elementary reactions represented by their stoichiometric
equations:
\begin{equation}\label{elementary reaction}
\sum_i\alpha_{\rho i}A_i \to \sum_i \beta_{\rho i} A_i \,
\end{equation}
The linear combinations $\sum_i\alpha_{\rho i}A_i$ and $\sum_i
\beta_{\rho i} A_i$ are the {\em complexes}. For each complex
$\sum_i y_{ji} A_i$ from the reaction mechanism we introduce an
intermediate auxiliary state, a {\em compound} $B_j$. Each
elementary reaction is represented in the form of the ``$2n$-tail
scheme'' (Figure~\ref{n-tail}) with two intermediate compounds:
\begin{equation}\label{stoichiometricequationcompaundI}
\sum_i\alpha_{\rho i}A_i \rightleftharpoons B_{\rho}^- \to
B_{\rho}^+ \rightleftharpoons \sum_i \beta_{\rho i} A_i
\,\vspace{-6pt}
\end{equation}

There are two main assumptions in the
Michaelis-Menten-Stueckelberg asymptotic:
\begin{itemize}
\item{The compounds are in fast equilibrium with the
    corresponding input reagents (QE);}\vspace{-10pt}
\item{They exist in very small concentrations compared to
    other components (QSS).}
\end{itemize}
The smallness of the concentration of the compounds implies that
they (i) have the perfect thermodynamic functions (entropy, internal
energy and free energy) and (ii) the rates of the reactions $B_i \to
B_j$ are linear functions of their concentrations.

One of the most important benefits from this approach is the
exclusion of the nonlinear reaction kinetics: They are in fast
equilibrium and equilibrium is ruled by thermodynamics.

Under the given smallness assumptions, the reaction rates $r_{\rho}$
for the elementary reactions have a special form of the {\em
generalized mass action law} (see Equation
(\ref{GeneralReactionRate}) below):
$$
r_{\rho}=\varphi_{\rho}\exp(\alpha_{\rho}, \check{\mu} ) \,
$$
where $\varphi_{\rho}>0$ is the {\em kinetic factor} and
$\exp(\alpha_{\rho}, \check{\mu} )$ is the Boltzmann factor. Here
and further in the text $(\alpha_{\rho}, \check{\mu} )=\sum_i
\alpha_{\rho i} \check{\mu}_i$ is the standard inner product,
$\exp(\ , \ )$ is the exponential of the value of the inner product
and $\check{\mu}_i$ are chemical potentials $\mu$ divided on $RT$.

For the prefect chemical systems, $\check{\mu}_i=\ln (c_i/c_i^*)$,
where $c_i$ is the concentration of $A_i$ and $c_i^*>0$ are the
positive equilibrium concentrations. For different values of the
conservation laws there are different positive equilibria. The
positive equilibrium $c_i^*$ is one of them and it is not important
which one is it. At this point, $\check{\mu}_i=0$, hence, the
kinetic factor for the perfect systems is just the equilibrium value
of the rate of the elementary reaction at the equilibrium point
$c^*$: $\varphi_{\rho}=r_{\rho}(c^*)$.

The linear kinetics of the compound reactions $B_i \to B_j$ implies
the remarkable identity for the reaction rates, the complex balance
condition (Equation (\ref{0=1}) below)
$$\sum_{\rho}\varphi_{\rho}\exp(\check{\mu},\alpha_{\rho})=
\sum_{\rho}\varphi_{\rho}\exp(\check{\mu},\beta_{\rho}) \, $$ for
all admissible values of $\check{\mu}$ and given $\varphi$ which may
vary independently. For other and more convenient forms of this
condition see Equation (\ref{complexbalanceGENKIN}) in
Section~\ref{Sec:GenKinThermod}. The complex balance condition is
sufficient for the positivity of the entropy production (for
decrease of the free energy under isothermal isochoric conditions).
The general formula for the reaction rate together with the complex
balance conditions and the positivity of the entropy production form
the Michaelis-Menten-Stueckelberg theorem
(Section~\ref{Sec:TheTeorem}).

The detailed balance conditions (Equation (\ref{detailed balance})
below),
$$
\varphi_{\rho}^+=\varphi_{\rho}^-
$$
for all ${\rho}$, are more restrictive than the complex balance
conditions. For the perfect systems, the detailed balance condition
takes the standard form: $r_{\rho}^+(c^*)=r_{\rho}^-(c^*)$.

We study also some other, less restrictive sufficient conditions for
accordance between thermodynamics and kinetics. For example, we
demonstrate that the $G$-inequality (Equation (\ref{0<1}) below)
$$\sum_{\rho}\varphi_{\rho}\exp(\check{\mu},\alpha_{\rho})\geq
\sum_{\rho}\varphi_{\rho}\exp(\check{\mu},\beta_{\rho}) \, $$ is
sufficient for the entropy growth and, at the same time, weaker than
the condition of complex balance.

If the reaction rates have the form of the generalized mass
action law but do not satisfy the sufficient condition of the
positivity of the entropy production, the situation may be
improved by the deformation of the entropy via addition of a
linear function. Such a deformation is always possible for the
{\em zero deficiency systems}. Let $q$ be the number of
different complexes in the reaction mechanism, $d$ be the
number of the connected components in the digraph of the
transitions between compounds (vertices are compounds and edges
are reactions). To exclude some degenerated cases a hypothesis
of {\em weak reversibility} is accepted: For any two vertices
$B_i$ and $B_j$, the existence of an oriented path from $B_i$
to $B_j$ implies the existence of an oriented path from $B_j$
to $B_i$.

Deficiency of the system is \cite{Feinberg1972}
$$q-d-{\rm rank}\Gamma \geq 0$$
where $\Gamma=(\gamma_{ij})=(\beta_{ij}-\alpha_{ij})$ is the
{\em stoichiometric matrix}. If the system has zero deficiency
then the entropy production becomes positive after the
deformation of the entropy via addition of a linear function.
The {\em deficiency zero theorem} in this form is proved in
Section~\ref{Sec:ComplexBalanceDeformEnt}.

Interrelations between the Michaelis-Menten-Stueckelberg
asymptotic and the transition state theory (which is also referred
to as the ``activated-complex theory'', ``absolute-rate theory'',
and ``theory of absolute reaction rates'') are very intriguing. This
theory was developed in 1935 by Eyring \cite{Eyring1935} and by
Evans and Polanyi \cite{EvansPolanyi1935}.

Basic ideas behind the transition state theory are
\cite{LaidlerTweedale2007}:
\begin{itemize}
\item{The activated complexes are in a quasi-equilibrium
    with the reactant molecules;}\vspace{-10pt}
\item{Rates of the reactions are studied by studying the
    activated complexes at the saddle point of a potential
    energy surface.}
\end{itemize}

The similarity is obvious but in the
Michaelis-Menten-Stueckelberg asymptotic an elementary reaction
is represented by a couple of compounds with the Markov
kinetics of transitions between them versus one transition
state, which moves along the ``reaction coordinate'', in the
transition state theory. This is not exactly the same approach
(for example, the theory of absolute reaction rates uses the
detailed balance conditions and does not produce anything
similar to the complex balance).

Important technical tools for the analysis of the
Michaelis-Menten-Stueckelberg asymptotic are the theorem about
preservation of the entropy production in the QE approximation
(Section~\ref{General} and Appendix 1) and the Morimoto
$H$-theorem for the Markov chains (Appendix 2).

\section{QE and Preservation of Entropy Production \label{General}}

In this section we introduce informally the QE approximation
and the important theorem about the preservation of entropy
production in this approximation. In Appendix 1, this
approximation and the theorem are presented with more formal
details.

Let us consider a system in a domain $U$ of a real vector space $E$
given by differential equations
\begin{equation}\label{sys1}
\frac{\D x}{\D t}=F(x)
\end{equation}

The QE approximation for (\ref{sys1}) uses two basic entities:
Entropy and slow variables.

{\em Entropy} $S$ is an increasing concave Lyapunov function
for (\ref{sys1}) with non-degenerated Hessian $\partial^2
S/\partial x_i \partial x_j$:
\begin{equation}\label{SecondLaw}
\frac{\D S}{\D t}\geq 0  \,
\end{equation}
In this approach, the increase of entropy in time is exploited
(the Second Law in the form (\ref{SecondLaw})).

The {\em slow variables} $M$ are defined as some differentiable
functions of variables $x$: $M=m(x)$. Here we assume that these
functions are linear. More general nonlinear theory was developed in
\cite{GorKarQE2006,GorKarProjector2004} with applications to the
Boltzmann equation and polymer physics. Selection of the slow
variables implies a hypothesis about separation of fast and slow
motion. The slow variables (almost) do not change during the fast
motion. After some initial time, the fast variables with high
accuracy are functions of the slow variables: We can write $x\approx
x^*_{M}$.

The QE approximation defines the functions $x^*_{M}$ as solutions to
the following {\bf MaxEnt} optimization~problem:
\begin{equation}\label{MaxEnt}
S(x) \to \max \;\; \mbox{subject to} \;\; m(x)=M \,
\end{equation}
The reasoning behind this approximation is simple: During the fast
initial layer motion, entropy increases and $M$ almost does not
change. Therefore, it is natural to assume that $x^*_{M}$ is close
to the solution to the MaxEnt optimization problem (\ref{MaxEnt}).
Further $x^*_{M}$ denotes a solution to the MaxEnt problem.

A solution to (\ref{MaxEnt}), $x^*_{M}$, is the {\em QE state}, the
set of the QE states $x^*_{M}$, parameterized by the values of the
slow variables $M$ is the {\em QE manifold}, the corresponding value
of entropy
\begin{equation}\label{QEentropy}
S^*(M)=S(x^*_{M})
\end{equation}
is the {\em QE entropy} and the equation for the slow variables
\begin{equation}\label{QEequation}
\frac{\D M}{\D t}= m(F(x^*_M))
\end{equation}
represents the {\em QE dynamics}.

The crucial property of the QE dynamics is the {\em preservation of
entropy production.}

\vspace{2mm}\noindent{\bf Theorem about preservation of entropy
production.} {\it Let us calculate  $\D S^*(M)/ \D t$ at point $M$
according to the QE dynamics (\ref{QEequation}) and find $\D S(x)/
\D t$ at point $x=x^*_M$ due to the initial system (\ref{sys1}). The
results always coincide:
\begin{equation}\label{PreservationEntProd}
\frac{\D S^*(M)}{\D t}=\frac{\D S(x)}{\D t} \,
\end{equation}}

The left hand side in (\ref{PreservationEntProd})  is computed due
to the QE approximation (\ref{QEequation}) and the right hand side
corresponds to the initial system (\ref{sys1}). The sketch of the
proof is given in Appendix 1.

The preservation of the entropy production leads to the {\em
preservation of the type of dynamics}: If for the initial
system (\ref{sys1}) entropy production is non-negative, $\D
S/\D t \geq 0$, then for the QE approximation
(\ref{QEequation}) the production of the QE entropy is also
non-negative, $\D S^*/\D t \geq 0$.

In addition, if for the initial system $({\D S}/{\D t})|_x = 0$ if
and only if $F(x)=0$ then the same property holds in the QE
approximation.

\section{The Classics and the Classical Confusion \label{MMConfusion}}

\subsection{The Asymptotic of Fast Reactions}

It is difficult to find who introduced the QE approximation. It was
impossible before the works of Boltzmann and Gibbs, and it became
very well known after the works of Jaynes \cite{Jaynes1}.

Chemical kinetics has been a source for model reduction ideas for
decades. The ideas of QE appear there very naturally: Fast reactions
go to their equilibrium and, after that, remain almost equilibrium
all the time. The general formalization of this idea looks as
follows. The kinetic equation has the form
\begin{equation}\label{generalkinetics}
\frac{\D N}{\D t}= K_{sl}(N)+\frac{1}{\epsilon}K_{fs}(N)
\end{equation}
Here $N$ is the vector of composition with components $N_i>0$,
$K_{sl}$ corresponds to the slow reactions, $K_{fs}$ corresponds to
fast reaction and $\epsilon > 0 $ is a small number. The system of
fast reactions has the linear conservation laws $b_l(N)=\sum_j
b_{lj}N_j$: $b_l(K_{fs}(N))\equiv 0$.

The fast subsystem
$$\frac{\D N}{\D t}=K_{fs}(N)$$
tends to a stable positive equilibrium $N^*$ for any positive
initial state $N(0)$ and this equilibrium is a function of the
values of the linear conservation laws $b_l(N(0))$. In the plane
$b_l(N)=b_l(N(0))$ the equilibrium is asymptotically stable and
exponentially attractive.

Vector $b(N)=(b_l(N))$ is the vector of slow variables and the QE
approximation is
\begin{equation}\label{QEreducedkinetics}
\frac{\D b}{\D t}= b(K_{sl}(N^*(b)) \,
\end{equation}

In chemical kinetics, equilibria can be described by conditional
entropy maximum (or conditional extremum of other thermodynamic
potentials). Therefore, for these cases we can apply the formalism
of the quasiequilibrium approximation. The thermodynamic Lyapunov
functions serve as tools for stability analysis and for model
reduction \cite{Hangos2010}.

The QE approximation, the asymptotic of fast reactions, is well
known in chemical kinetics. Another very important approximation was
invented in chemical kinetics as well. It is the Quasi Steady State
(QSS) approximation. QSS was proposed in \cite{Bodenstein1913} and
was elaborated into an important tool for analysis of chemical
reaction mechanisms and kinetics
\cite{Semenov1939,Christiansen1953,Helfferich1989}. The classical
QSS is based on the {\em relative smallness of concentrations} of
some of ``active'' reagents (radicals, substrate-enzyme complexes or
active components on the catalyst surface) \cite{Aris1965,Segel89}.
In the enzyme kinetics, its invention was traditionally connected to
the so-called Michaelis-Menten kinetics.

\subsection{QSS and the Briggs-Haldane Asymptotic}

Perhaps the first very clear explanation of the QSS was given
by Briggs and Haldane in 1925 \cite{BriggsHaldane1925}. Briggs
and Haldane consider the simplest enzyme reaction $S+E
\leftrightharpoons SE \to P+E$ and mention that the total
concentration of enzyme ($[E]+[SE]$) is ``negligibly small''
compared with the concentration of substrate $[S]$. After that
they conclude that $\frac{\D}{\D t}{[SE]}$ is ``negligible''
compared with $\frac{\D}{\D t}{[S]}$ and $\frac{\D}{\D t}{[P]}$
and produce the now famous  `Michaelis-Menten' formula, which
was unknown to Michaelis and Menten: $k_1
[E][S]=(k_{-1}+k_2)[ES]$ or
\begin{equation}\label{MichaelisMentenFormula}
[ES]=\frac{[E][S]}{K_M+[S]} \; \; {\rm and} \; \; \frac{\D}{\D
t}{[P]}=k_2 [ES]=\frac{k_2[E][S]}{K_M+[S]}\,
\end{equation}
where  the ``Michaelis-Menten constant'' is
$$K_M=\frac{k_{-1}+k_2}{k_1} $$

There is plenty of misleading comments in later publications about
QSS. Two most important confusions are:
\begin{itemize}
\item{Enzymes (or catalysts, or radicals) participate in
    {\em fast reactions} and, hence, relax faster than
    substrates or stable components. This is obviously
    wrong for many QSS systems: For example, in the
    Michaelis-Menten system {\em all} reactions include
    enzyme together with substrate or product. There are no
    separate fast reactions for enzyme without substrate or
    product.}\vspace{-10pt}
\item{{\em Concentrations of intermediates are constant}
    because in QSS we equate their time derivatives to
    zero. In general case, this is also wrong: We equate
    the kinetic expressions for some time derivatives to
    zero, indeed, but this just exploits the fact that the
    time derivatives of intermediates concentrations are
    small together with their values, but not obligatory
    zero. If we accept QSS then these derivatives are not
    zero as well: To prove this we can just differentiate
    the Michaelis-Menten formula
    (\ref{MichaelisMentenFormula}) and find that [ES] in
    QSS is almost constant when $[S]\gg K_M$, this is an
    additional condition, different from the
    Briggs-Haldane condition $[E]+[AE]\ll [S]$ (for more
    details \linebreak see \cite{Segel89,Klonowski1983,Yablonskii1991}
    and a simple detailed case study \cite{LiShenLi2008}).}
\end{itemize}

After a simple transformation of variables the QSS smallness of
concentration transforms into a separation of time scales in a
standard singular perturbation form (see, for example
\cite{Yablonskii1991,GorbKarlinChemEngS2003}). Let us demonstrate
this on the traditional Michaelis-Menten system:
\begin{equation}\label{tradMMsystem}
\begin{split}
&\frac{\D [S]}{\D t}=- k_1 [S][E]+k_{-1}[SE] \,  \\
&\frac{\D [SE]}{\D t}= k_1 [S][E]-(k_{-1}+k_2)[SE] \, \\
&[E]+[SE]=e=const, \, [S]+[P]=s=const \,
\end{split}
\end{equation}
This is a homogeneous system with the isochoric (fixed volume)
conditions for which we write the equations. The Briggs-Haldane
condition is $e\ll s$. Let us use dimensionless variables
$x=[S]/s$, $\xi=[SE]/e$:
\begin{equation}
\begin{split}
&\frac{s}{e}\frac{\D x}{\D t}=- sk_1 x(1-\xi)+k_{-1}\xi \,  \\
&\frac{\D \xi}{\D t}= sk_1 x(1-\xi)-(k_{-1}+k_2)\xi \,
\end{split}
\end{equation}

To obtain the standard singularly perturbed system with the small
parameter at the derivative, we need to change the time scale. This
means that when $e \to 0$ the reaction goes proportionally slower
and to study this limit properly we have to adjust the time scale:
$\D \tau =\frac{e}{s}\D t$:
\begin{equation}\label{SingPertMM}
\begin{split}
&\frac{\D x}{\D \tau}=- sk_1 x(1-\xi)+k_{-1}\xi \,  \\
&\frac{e}{s}\frac{\D \xi}{\D \tau}= sk_1 x(1-\xi)-(k_{-1}+k_2)\xi \,
\end{split}
\end{equation}
For small $e/s$, the second equation is a fast subsystem. According
to this fast equation, for a given constant $x$, the variable $\xi$
relaxes to $$\xi_{\rm QSS}=\frac{s x}{K_M+sx}$$ exponentially, as
$\exp(-(sk_1x+k_{-1}+k_2)t)$. Therefore, the classical singular
perturbation theory based on the Tikhonov theorem
\cite{Tikhonov1952,Vasil'eva1963} can be applied to the system in
the form (\ref{SingPertMM}) and the QSS approximation is applicable
even on an infinite time interval \cite{Hoppensteadt1966}. This
transformation of variables and introduction of slow time is a
standard procedure for rigorous proof of QSS validity in catalysis
\cite{Yablonskii1991}, enzyme kinetics~\cite{Battelu1986} and other
areas of kinetics and chemical engineering \cite{Aris1965}.

It is worth to mention that the smallness of parameter $e/s$ can be
easily controlled in experiments, whereas the time derivatives,
transformation rates and many other quantities just appear as a
result of kinetics and cannot be controlled directly.

\subsection{The Michaelis and Menten Asymptotic}

QSS is not QE but the classical work of Michaelis and Menten
\cite{MichaelisMenten1913} was done on the intersection of QSS and
QE. After the brilliantly clear work of Briggs and Haldane, the name
``Michaelis-Menten'' was attached to the Briggs and Haldane
equation and the original work of Michaelis and Menten was
considered as an important particular case of this approach, an
approximation with additional and not necessary assumptions of QE.
From our point of view, the Michaelis-Menten work includes more and
may give rise to an important general class of kinetic models.

Michaelis and Menten studied the ``fermentative splitting of
cane sugar''. They introduced three ``compounds'': The
sucrose-ferment combination, the fructose-ferment combination
and the glucose-ferment combination. The fundamental assumption
of their work was ``that the rate of breakdown at any moment is
proportional to the concentration of the sucrose-invertase
compound''.

They started from the assumption that at any moment according to the
mass action law
\begin{equation}\label{MMQEass}
[S_i][E]=K_i [S_iE]
\end{equation}
where $[S_i]$ is the concentration of the $i$th sugar (here,
$i=0$ for sucrose, 1 for fructose and 2 for glucose), $[E]$ is
the concentration of the free invertase and $K_i$ is the $i$th
equilibrium constant.

For simplification, they use the assumption that the concentration
of any sugar in question in {\em free state} is practically equal to
that of the total sugar in question.

Finally, they obtain
\begin{equation}\label{properMM}
[S_0E]=\frac{e [S_0]}{K_0(1+q[P])+[S_0]}\,
\end{equation}
where $e=[E]+\sum_i [S_iE]$, $[P]=[S_1]=[S_2]$ and
$q=\frac{1}{K_1}+\frac{1}{K_2}$.

Of course, this formula may be considered as a particular case of
the Briggs-Haldane formula (\ref{MichaelisMentenFormula}) if we
take $k_{-1}\gg k_2$ in (\ref{MichaelisMentenFormula}) ({\em i.e.},
the equilibration $S+E\leftrightharpoons SE$ is much faster than the
reaction $SE \to P+E$) and assume that $q=0$ in (\ref{properMM})
({\em i.e.},  fructose-ferment combination and glucose-ferment
combination are practically absent).

This is the truth but may be not the complete truth. The
Michaelis-Menten approach with many compounds which are present in
small amounts and satisfy the QE assumption (\ref{MMQEass}) is a
seed of the general kinetic theory for perfect and non-perfect
mixtures.

\section{Chemical Kinetics and QE Approximation
\label{ChemKinQEApprox}}

\subsection{Stoichiometric Algebra and Kinetic Equations}

In this section, we introduce the basic notations of the chemical
kinetics formalism. For more details see, for example,
\cite{Yablonskii1991}.

The list of components is a finite set of symbols $A_1, \ldots,
A_n$.

A reaction mechanism is a finite set of the {\em stoichiometric
equations} of elementary reactions:
\begin{equation}\label{stoichiometricequation}
\sum_i\alpha_{\rho i}A_i \to \sum_i \beta_{\rho i} A_i \,
\end{equation}
where $\rho =1, \ldots, m$ is the reaction number and the {\em
stoichiometric coefficients} $\alpha_{\rho i},\beta_{\rho i}$ are
nonnegative~integers.

A {\em stoichiometric vector} $\gamma_{\rho}$ of the reaction
(\ref{stoichiometricequation}) is a $n$-dimensional vector with
coordinates
\begin{equation}
\gamma_{\rho i}=\beta_{\rho i}-\alpha_{\rho i} \,
\end{equation}
that is, ``gain minus loss'' in the $\rho$th elementary reaction.

A nonnegative extensive variable $N_i$, the amount of $A_i$,
corresponds to each component. We call the vector $N$ with
coordinates $N_i$ ``the composition vector''. The concentration of
$A_i$ is an intensive variable $c_i=N_i/V$, where $V>0$ is the
volume. The vector $c=N/V$ with coordinates $c_i$ is the vector of
concentrations.

A non-negative intensive quantity, $r_{\rho}$, the reaction rate,
corresponds to each reaction (\ref{stoichiometricequation}). The
kinetic equations in the absence of external fluxes are
\begin{equation}\label{KinUrChem}
\frac{\D N}{\D t}=V \sum_{\rho}r_{\rho} \gamma_{\rho}
\end{equation}
If the volume is not constant then equations for concentrations
include $\dot{V}$ and have different form (this is typical for the
combustion reactions, for example).

For perfect systems and not so fast reactions, the reaction rates
are functions of concentrations and temperature given by the {\em
mass action law} for the dependance on concentrations and by the
generalized Arrhenius equation for the dependance on temperature
$T$.

The mass action law states:
\begin{equation}\label{MAL}
r_{\rho}(c,T)=k_{\rho}(T)\prod_i c_i^{\alpha_{\rho i}} \,
\end{equation}
where $k_{\rho}(T)$ is the reaction rate constant.

The generalized Arrhenius equation is:
\begin{equation}\label{Arrhenius}
k_{\rho}(T)=A_{\rho}\exp\left(\frac{S_{{\rm
a}\rho}}{R}\right)\exp \left(-\frac{E_{{\rm a}\rho}}{RT}\right) \,
\end{equation}
where $R=8.314\,472~\frac{\mathrm{J}}{\mathrm{K~mol}}$ is the
universal, or ideal gas constant, $E_{{\rm a}\rho}$ is the
activation energy, $S_{{\rm a}\rho}$ is the activation entropy ({\em
i.e.}, $E_{{\rm a}\rho}-TS_{{\rm a}\rho}$ is the activation free
energy), $A_{\rho}$ is the constant pre-exponential factor. Some
authors neglect the $S_{{\rm a}\rho}$ term because it may be less
important than the activation energy, but it is necessary to stress
that without this term it may be impossible to reconcile the kinetic
equations with the classical thermodynamics.

In general, the constants for different reactions are not
independent. They are connected by various conditions that follow
from thermodynamics (the second law, the entropy growth for isolated
systems) or microreversibility assumption (the detailed balance and
the Onsager reciprocal relations). In Section~\ref{Sec:Accordance}
we discuss these conditions in more general settings.

For nonideal systems, more general kinetic law is needed. In
Section~\ref{GeneralKinetics} we produce such a general law
following the ideas of the original Michaelis and Menten paper (this
is not the same as the famous ``Michaelis-Menten kinetics''). For
this work we need a general formalism of QE approximation for
chemical kinetics.

\subsection{Formalism of QE Approximation for Chemical Kinetics}

\subsubsection{4.2.1. QE Manifold}

In this section, we describe the general formalism of the QE for
chemical kinetics following \cite{GorbKarlinChemEngS2003}.

The general construction of the quasi-equilibrium manifold gives the
following procedure. First, let us consider the chemical reactions
in a constant volume under the isothermal conditions. The free
energy $F(N,T)=Vf(c,T)$ should decrease due to reactions. In the
space of concentrations, one defines a subspace of fast motions $L$.
It should be spanned by the stoichiometric vectors of {\em fast
reactions}.

Slow coordinates are linear functions that annulate $L$. These
functions form a subspace in the space of linear functions on the
concentration space. Dimension of this space is $s=n-\dim L$. It is
necessary to choose any basis in this subspace. We can use for this
purpose a basis $b_j$ in $L^{\perp}$, an orthogonal complement to
$L$ and define the basic functionals as $b_j(N)=(b_j,N)$.

The description of the QE manifold is very simple in the Legendre
transform. The chemical potentials are partial derivatives
\begin{equation}\label{chemical potential}
\mu_i=\frac{\partial F(N,T)}{\partial N_i}=\frac{\partial
f(c,T)}{\partial c_i}\,
\end{equation}
Let us use $\mu_i$ as new coordinates. In these new coordinates (the
``conjugated coordinates''), the QE manifold is just an orthogonal
complement to $L$. This subspace, $L^{\perp}$, is defined by
equations
\begin{equation}\label{Lperpendicular}
\sum_i \mu_i \gamma_i = 0 \;\; {\rm for \; any}\;\; \gamma \in L
\end{equation}
It is sufficient to take in (\ref{Lperpendicular}) not all $\gamma
\in L$ but only elements from a basis in $L$. In this case, we get
the system of $n-\dim L$ linear equations of the form
(\ref{Lperpendicular}) and their solution does not cause any
difficulty. For the actual computations, one requires the inversion
from $\mu$ to $c$.

It is worth to mention that the problems of the selection of the
slow variables and of the description of the QE manifold in the
conjugated variables can be considered as the same problem of
description of the orthogonal complement, $L^{\perp}$.

To finalize the construction of the QE approximation, we should find
for any given values of slow variables (and of conservation laws)
$b_i$ the corresponding point on the QE manifold. This means that we
have to solve the system of equations for $c$:
\begin{equation}\label{EQforQE}
b(N)=b;\;\; (\mu (c,T),\gamma_{\rho})=0\,
\end{equation}
where $b$ is the vector of slow variables, $\mu$ is the vector of
chemical potentials and vectors $\gamma_{\rho}$ form a basis in $L$.
After that, we have the QE dependence $c_{QE}(b)$ and for any
admissible value of $b$ we can find all the reaction rates and
calculate $\dot{b}$.

Unfortunately, the system (\ref{EQforQE}) can be solved analytically
only in some special cases. In general case, we have to solve it
numerically. For this purpose, it may be convenient to keep the
optimization statement of the problem: $F \to \min$ subject to given
$b$. There exists plenty of methods of convex optimization for
solution of this problem.

The standard toy example gives us a fast dissociation reaction. Let a homogeneous
reaction mechanism include a fast reaction of the form $A+B \rightleftharpoons AB$.  We
can easily find the QE approximation for this fast reaction. The slow variables are the
quantities $b_1=N_A-N_B$ and $b_2=N_A+N_{AB}$ which do not change in this reaction. Let
the chemical potentials be $\mu_A/RT=\ln c_A + \mu_{A0}$, $\mu_B/RT=\ln c_B + \mu_{B0}$,
$\mu_{AB}/RT=\ln c_{AB} + \mu_{AB0}$. This corresponds to the free energy $F=VRT \sum_i
c_i (\ln c_i +\mu_{i0})$, the correspondent free entropy (the Massieu-Planck potential)
is $-F/T$. The stoichiometric vector is $\gamma=(-1,-1,1)$ and the equations
(\ref{EQforQE}) take the form
$$c_A-c_B=\frac{b_1}{V}\, , \;\; c_A+c_{AB}=\frac{b_2}{V}\,
, \;\; \frac{c_{AB}}{c_Ac_B}=K\, $$
 where $K$ is the equilibrium constant
 $K=\exp(\mu_{A0}+\mu_{B0}-\mu_{AB0})$.

From these equations we get the expressions for the QE
concentrations:
 $$c_A(b_1,b_2)=\frac{b_1}{2V}-\frac{1}{2K}+\sqrt{\left(\frac{b_1}{2V}-\frac{1}{2K}\right)^2+\frac{b_2}{KV}
 }$$
 $$c_B(b_1,b_2)=c_A(b_1,b_2)-\frac{b_1}{V}\, , \;\;
c_{AB}(b_1,b_2)=\frac{b_2}{V}-c_A(b_1,b_2)$$

The QE free entropy is the value of the free entropy at this point,
$c(b_1,b_2)$.

\subsubsection{4.2.2. QE in Traditional MM System}

Let us return to the simplest homogeneous enzyme reaction $E+S \rightleftharpoons ES
\rightarrow P+E$, the traditional Michaelis-Menten System (\ref{tradMMsystem}) (it is
simpler than the system studied by Michaelis and Menten \cite{MichaelisMenten1913}). Let
us assume that the reaction $E+S \rightleftharpoons ES$ is fast. This means that both
$k_1$ and $k_{-1} $ include large parameters: $k_1=\frac{1}{\epsilon}\kappa_1$,
$k_{-1}=\frac{1}{\epsilon}\kappa_{-1}$. For small $\epsilon$, we will apply the QE
approximation. Only three components participate in the fast reaction, $A_1=S$, $A_2=E$,
$A_3=ES$. For analysis of the QE manifold we do not need to involve other components.

The stoichiometric vector of the fast reaction is
$\gamma=(-1,-1,1)$. The space $L$ is one-dimensional and its basis
is this vector $\gamma$. The space $L^{\perp}$ is two-dimensional
and one of the convenient bases is $b_1=(1,0,1)$, $b_2=(0,1,1)$. The
corresponding slow variables are $b_1(N)=N_1+N_3$, $b_2(N)=N_2+N_3$.
The first slow variable is the sum of the free substrate and the
substrate captured in the enzyme-substrate complex. The second of
them is the conserved quantity, the total amount of enzyme.

The equation for the QE manifold is (\ref{MMQEass}): $k_1 c_1
c_2=k_{-1}c_3$ or
$\frac{c_1}{c_1^*}\frac{c_2}{c_2^*}=\frac{c_3}{c_3^*}$ because $k_1
c_1^*c_2^*=k_{-1}c_3^*$, where $c_i^*=c_i^*(T)>0$ are the so-called
standard equilibrium values and for perfect systems \linebreak
$\mu_i=RT\ln(c_i/c_i^*)$, $F=RTV\sum_i c_i (\ln(c_i/c_i^*)-1)$.

Let us fix the slow variables and find $c_{1,2,3}$. Equations
(\ref{EQforQE}) turn into
$$c_1+c_3=b_1\, , \; c_2+c_3=b_2\, , \; k_1 c_1
c_2=k_{-1}c_3 $$ Here we change dynamic variables from $N$ to $c$
because this is a homogeneous system with constant~volume.

If we use $c_1=b_1-c_3$ and $c_2=b_2-c_3$ then we obtain a quadratic
equation for $c_3$:
\begin{equation}\label{quadraticMMeq}
k_1c_3^2-(k_1b_1+k_1b_2+k_{-1})c_3+k_1b_1b_2=0\,
\end{equation}
Therefore,
$$c_3(b_1,b_2)=\frac{1}{2}\left(b_1+b_2+\frac{k_{-1}}{k_1}\right)-
\frac{1}{2}\sqrt{\left(b_1+b_2+\frac{k_{-1}}{k_1}\right)^2-4b_1b_2}
$$ The sign ``$-$'' is selected to provide positivity of all $c_i$.
This choice provides also the proper asymptotic: $c_3\to 0$ if any
of $b_i\to 0$. For other $c_{1,2}$ we should use $c_1=b_1-c_3$ and
$c_2=b_2-c_3$.

The time derivatives of concentrations are:
\begin{equation}
\begin{split}
&\dot{c}_1=-k_1 c_1c_2+k_{-1}c_3+v_{\rm in} c_1^{\rm in} -v_{\rm out}c_1 \, \\
&\dot{c}_2=-k_1 c_1c_2+(k_{-1}+k_2)c_3+v_{\rm in} c_2^{\rm in} -v_{\rm out}c_2 \,  \\
&\dot{c}_3=k_1 c_1c_2-(k_{-1}+k_2)c_3+v_{\rm in} c_3^{\rm in}-
v_{\rm out}c_3 \,  \\
&\dot{c}_4=k_2c_3 + v_{\rm in} c_4^{\rm in}- v_{\rm out}c_4 \,
\end{split}
\end{equation}
here we added external flux with input and output velocities (per
unite volume) $v_{\rm in}$ and $v_{\rm out}$ and input
concentrations $c^{\rm in}$. This is done to stress that the QE
approximation holds also for a system with fluxes if the fast
equilibrium subsystem is fast enough. The input and output
velocities are the same for all components because the system is
homogeneous.

The slow system is
\begin{equation}
\begin{split}
&\dot{b}_1=\dot{c}_1+\dot{c}_3=-k_2c_3+v_{\rm in} b_1^{\rm in} -v_{\rm out}b_1 \,  \\
&\dot{b}_2=\dot{c}_2+\dot{c}_3=v_{\rm in} b_2^{\rm in} -v_{\rm out}b_2 \,  \\
&\dot{c}_4=k_2c_3 + v_{\rm in} c_4^{\rm in}- v_{\rm out}c_4 \,
\end{split}
\end{equation}
where $b_1^{\rm in}=c_1^{\rm in}+c_3^{\rm in}$, $b_2^{\rm
in}=c_2^{\rm in}+c_3^{\rm in}$.

Now, we should use the expression for $c_3(b_1,b_2)$:
\begin{equation}\label{MMreducedQE}
\begin{split}
\dot{b}_1=&-k_2\frac{1}{2}\left[\left(b_1+b_2+\frac{k_{-1}}{k_1}\right)-
\frac{1}{2}\sqrt{\left(b_1+b_2+\frac{k_{-1}}{k_1}\right)^2-4b_1b_2}\,\right]+v_{\rm in} b_1^{\rm in} -v_{\rm out}b_1 \,  \\
\dot{c}_4=&k_2\frac{1}{2}\left[\left(b_1+b_2+\frac{k_{-1}}{k_1}\right)-
\frac{1}{2}\sqrt{\left(b_1+b_2+\frac{k_{-1}}{k_1}\right)^2-4b_1b_2}\,\right]+ v_{\rm in} c_4^{\rm in}- v_{\rm out}c_4 \,  \\
\dot{b}_2=&v_{\rm in} b_2^{\rm in} -v_{\rm out}b_2 \,
\end{split}
\end{equation}
It is obvious here that in the reduced system (\ref{MMreducedQE})
there exists one reaction from the lumped component with
concentration $b_1$ (the total amount of substrate in free state and
in the substrate-enzyme complex) into the component (product) with
concentration $c_4$. The rate of this reaction is $k_2c(b_1b_2)$.
The lumped component with concentration $b_2$ (the total amount of
the enzyme in free state and in the substrate-enzyme complex)
affects the reaction rate but does not change in the reaction.

Let us use for simplification of this system the assumption of the
substrate excess (we follow the logic of the original Michaelis and
Menten paper \cite{MichaelisMenten1913}):
\begin{equation}\label{substrateExcess}
[S]\gg [SE]\,, \;\; {\it i.e.}, \;\; b_1 \gg c_3 \,
\end{equation}
Under this assumption, the quadratic equation (\ref{quadraticMMeq})
transforms into
\begin{equation}\label{linearMMeq}
\left(1+\frac{b_2}{b_1}+\frac{k_{-1}}{k_1b_1}\right)c_3=b_2+o\left(\frac{c_3}{b_1}\right)\,
\end{equation}
and in this approximation
\begin{equation}\label{MMlinear}
c_3=\frac{b_2b_1}{b_1+b_2+\frac{k_{-1}}{k_1}}
\end{equation}
(compare to (\ref{properMM}) and (\ref{MichaelisMentenFormula}):
This equation includes an additional term $b_2$ in denominator
because we did not assume formally anything about the smallness of
$b_2$ in (\ref{substrateExcess})).

After this simplification, the QE slow equations (\ref{MMreducedQE})
take the form
\begin{equation}
\begin{split}
&\dot{b}_1=-\frac{k_2b_2b_1}{b_1+b_2+\frac{k_{-1}}{k_1}}+v_{\rm in} b_1^{\rm in} -v_{\rm out}b_1 \, \\
&\dot{b}_2=v_{\rm in} b_2^{\rm in} -v_{\rm out}b_2 \,  \\
&\dot{c}_4=\frac{k_2b_2b_1}{b_1+b_2+\frac{k_{-1}}{k_1}} + v_{\rm in}
c_4^{\rm in}- v_{\rm out}c_4 \,
\end{split}
\end{equation}
This is the typical form in the reduced equations for catalytic
reactions: Nominator in the reaction rate corresponds to the
``brutto reaction'' $S+E\to P+E$ \cite{Yablonskii1991,YAbLaz1997}.

\subsubsection{4.2.3. Heterogeneous Catalytic Reaction}

For the second example, let us assume equilibrium with respect to
the adsorption in the CO on Pt oxidation:
$$\mbox{CO+Pt$\rightleftharpoons$PtCO;
O$_2$+2Pt$\rightleftharpoons$2PtO}$$ (for detailed discussion of the
modeling of CO on Pt oxidation, this ``Mona Liza'' of catalysis, we
address readers to \cite{Yablonskii1991}). The list of components
involved in these 2 reactions is: $A_1=$ CO, $A_2=$ O$_2$, $A_3=$
Pt, $A_4=$ PtO, $A_5=$ PtCO (CO$_2$ does not participate in
adsorption and may be excluded at this point).

Subspace $L$ is two-dimensional. It is spanned by the stoichiometric
vectors, $\gamma_1=(-1,0,-1,0,1)$, $\gamma_2=(0,-1,-2,2,0)$.

The orthogonal complement to $L$ is a three-dimensional subspace
spanned by vectors $(0,2,0,1,0)$, $(1,0,0,0,1)$, $(0,0,1,1,1)$. This
basis is not orthonormal but convenient because of integer
coordinates.

The corresponding slow variables are
\begin{equation}\label{slowCO2}
\begin{split}
&b_1=2N_2+N_4=2N_{\rm O_2}+N_{\rm PtO}\,  \\
&b_2=N_1+N_5=N_{\rm CO} + N_{\rm PtCO}\, \\
&b_3=N_3+N_4+N_5=N_{\rm Pt}+N_{\rm PtO}+N_{\rm PtCO}\,
\end{split}
\end{equation}
For heterogeneous systems, caution is needed in transition between
$N$ and $c$ variables because there are two ``volumes'' and we
cannot put in (\ref{slowCO2}) $c_i$ instead of $N_i$: $N_{\rm
gas}=V_{\rm gas}c_{\rm gas}$ but $N_{\rm surf}=V_{\rm surf}c_{\rm
surf}$, where where $V_{\rm gas}$ is the volume of gas, $V_{\rm
surf}$ is the area of surface.

There is a law of conservation of the catalyst: $N_{\rm Pt}+N_{\rm
PtO}+N_{\rm PtCO}=b_3=const$. Therefore, we have two non-trivial
dynamical slow variables, $b_1$ and $b_2$. They have a very clear
sense: $b_1$ is the amount of atoms of oxygen accumulated in O$_2$
and PtO and $b_2$ is the amount of atoms of carbon accumulated in CO
and PtCO.

The free energy for the perfect heterogeneous system has the form
\begin{equation}\label{FreeEnergyPerfectHeter}
F=V_{\rm gas}RT\sum_{A_i\,{\rm  gas}} c_i
\left(\ln\left(\frac{c_i}{c_i^*}\right)-1\right)+ V_{\rm
surf}RT\sum_{A_i\,{\rm  surf}} c_i
\left(\ln\left(\frac{c_i}{c_i^*}\right)-1\right)\,
\end{equation}
where $c_i$ are the corresponding concentrations and
$c_i^*=c_i^*(T)>0$ are the so-called standard equilibrium values.
(The QE free energy is the value of the free energy at the QE
point.)

From the expression (\ref{FreeEnergyPerfectHeter}) we get the
chemical potentials of the perfect mixture
\begin{equation}\label{perfectPotential}
\mu_i=RT\ln\left(\frac{c_i}{c_i^*}\right) \,
\end{equation}

The QE manifold in the conjugated variables is given by equations:
$$-\mu_1-\mu_3+\mu_5=0\,;\; -\mu_2-2\mu_3+2\mu_4=0 $$
It is trivial to resolve these equations with respect to
$\mu_{3,4}$, for example:
$$\mu_4=\frac{1}{2}\mu_2+\mu_3\, ; \; \mu_5=\mu_1+\mu_3$$
or with the standard equilibria:
$$
\frac{c_4}{c_4^*}=\frac{c_3}{c_3^*}\sqrt{\frac{c_2}{c_2^*}}\, , \;
\frac{c_5}{c_5^*}=\frac{c_1}{c_1^*}\frac{c_3}{c_3^*}
$$
or in the kinetic form (we assume that the kinetic constants are in
accordance with thermodynamics and all these forms are equivalent):
\begin{equation}\label{QECOKIN}
k_1 c_1c_3=k_{-1}c_5\,, \; k_2 c_2c_3^2=k_{-2} c_4^2 \,
\end{equation}

The next task is to solve the system of equations:
\begin{equation}\label{slowCO2qeeq}
\begin{split}
&k_1 c_1c_3=k_{-1}c_5\,, \; k_2 c_2c_3^2=k_{-2} c_4^2 \,,2 V_{\rm
gas}c_2+V_{\rm surf}c_4=b_1\, , \; \\& V_{\rm gas}c_1+V_{\rm
surf}c_5=b_2\,, \; V_{\rm surf}(c_3+c_4+c_5)=b_3
\end{split}
\end{equation}
This is a system of five equations with respect to five unknown
variables, $c_{1,2,3,4,5}$. We have to solve them and use the
solution for calculation of reaction rates in the QE equations for
the slow variables. Let us construct these equations first, and then
return to (\ref{slowCO2qeeq}).

We assume the adsorption (the Langmuir-Hinshelwood) mechanism of CO
oxidation (the numbers in parentheses are used below for the
numeration of the reaction rate constants):
\begin{equation}\label{Langmuir-Hinshelwood}
\begin{split}
(\pm 1)\;\mbox{CO+Pt$\rightleftharpoons$PtCO}\, \\
(\pm 2)\; \mbox{O$_2$+2Pt$\rightleftharpoons$2PtO}\,  \\
(3)\; \mbox{PtO+PtCO$\to$CO$_2$+2Pt} \;
\end{split}
\end{equation}

The kinetic equations for this system (including the flux in the gas
phase) is
\begin{eqnarray}\label{COoxidEQ}
&{\rm CO}\;&\dot{N}_1=V_{\rm surf}(-k_1 c_1c_3+k_{-1} c_5) + V_{\rm gas}(v_{\rm in} c_1^{\rm in}- v_{\rm out}c_1 ) \,  \nonumber\\
&{\rm O_2}&\dot{N}_2=V_{\rm surf}(-k_2 c_2 c_3^2+ k_{-2}c_4^2)+ V_{\rm gas}(v_{\rm in} c_2^{\rm in}- v_{\rm out}c_2 ) \, \nonumber\\
&{\rm Pt}&\dot{N}_3=V_{\rm surf} (-k_1 c_1c_3+k_{-1} c_5 -2k_2 c_2 c_3^2+ 2k_{-2}c_4^2 +2k_3c_4c_5)\, \\
&{\rm PtO}&\dot{N}_4 =V_{\rm surf}(2k_2 c_2 c_3^2- 2k_{-2}c_4^2 - k_3c_4c_5)\, \nonumber\\
&{\rm PtCO}\;&\dot{N}_5 =V_{\rm surf} (k_1 c_1c_3-k_{-1} c_5- k_3c_4c_5)\, \nonumber\\
&{\rm CO_2}&\dot{N}_6=V_{\rm surf}k_3c_4c_5+V_{\rm gas}( v_{\rm in}
c_6^{\rm in}- v_{\rm out}c_6)\, \nonumber
\end{eqnarray}
Here $v_{\rm in}$ and $v_{\rm out}$ are the flux rates (per unit
volume).

For the slow variables this equation gives:
\begin{equation}\label{slowCOQE eq}
\begin{split}
&\dot{b}_1=2\dot{N}_2+\dot{N}_4=-V_{\rm surf}k_3c_4c_5+ 2V_{\rm gas}(v_{\rm in} c_2^{\rm in}- v_{\rm out}c_2 )\\
&\dot{b}_2=\dot{N}_1+\dot{N}_5=-V_{\rm surf}k_3c_4c_5+ V_{\rm gas}(v_{\rm in} c_1^{\rm in}- v_{\rm out}c_1 )\\
&\dot{b}_3=\dot{N}_3+\dot{N}_4+\dot{N}_5=0 \\
&\dot{N}_6=V_{\rm surf}k_3c_4c_5+V_{\rm gas}( v_{\rm in} c_6^{\rm
in}- v_{\rm out}c_6)
\end{split}
\end{equation}
This system looks quite simple. Only one reaction,
\begin{equation}\label{COinteraction}
\mbox{PtO+PtCO$\to$CO$_2$+2Pt}\,
\end{equation}
is visible. If we know expressions for $c_{3,5}(b)$ then this
reaction rate is also known. In addition, to work with the rates of
fluxes, the expressions for $c_{1,2}(b)$ are needed.

The system of equations (\ref{slowCO2qeeq}) is explicitly solvable
but the result is quite cumbersome. Therefore, let us consider its
simplification without explicit analytic solution. We assume the
following smallness:
\begin{equation}\label{COsmallness}
b_1 \gg N_4\, , \;\; b_2 \gg N_5\,
\end{equation}
Together with this smallness assumptions equations
(\ref{slowCO2qeeq}) give:
\begin{equation}
\begin{split}
&c_3=\frac{b_3}{V_{\rm
surf}\left(1+\frac{k_1}{k_{-1}}\frac{b_2}{V_{\rm
gas}}+\sqrt{\frac{1}{2}\frac{k_2}{k_{-2}}\frac{b_1}{V_{\rm gas}}}\right)} \\
&c_4=\sqrt{\frac{1}{2}\frac{k_2}{k_{-2}}\frac{b_1}{V_{\rm
gas}}}\frac{b_3}{V_{\rm
surf}\left(1+\frac{k_1}{k_{-1}}\frac{b_2}{V_{\rm
gas}}+\sqrt{\frac{1}{2}\frac{k_2}{k_{-2}}\frac{b_1}{V_{\rm gas}}}\right)}  \\
&c_5=\frac{k_1}{k_{-1}}\frac{b_2}{V_{\rm gas}}\frac{b_3}{V_{\rm
surf}\left(1+\frac{k_1}{k_{-1}}\frac{b_2}{V_{\rm
gas}}+\sqrt{\frac{1}{2}\frac{k_2}{k_{-2}}\frac{b_1}{V_{\rm
gas}}}\right)}
\end{split}
\end{equation}
In this approximation, we have for the reaction
(\ref{COinteraction}) rate
$$r=k_3 c_4 c_5 =k_3 \frac{k_1}{k_{-1}}\sqrt{\frac{1}{2}\frac{k_2}{k_{-2}}} \frac{\sqrt{b_1}b_2}{V_{\rm
gas}^{3/2}} \frac{b_3^2}{V_{\rm
surf}^2\left(1+\frac{k_1}{k_{-1}}\frac{b_2}{V_{\rm
gas}}+\sqrt{\frac{1}{2}\frac{k_2}{k_{-2}}\frac{b_1}{V_{\rm
gas}}}\right)^2}
$$
This expression gives the closure for the slow QE equations
(\ref{slowCOQE eq}).

\subsubsection{4.2.3. Discussion of the QE procedure for Chemical Kinetics}

We finalize here the illustration of the general QE procedure for
chemical kinetics. As we can see, the simple analytic description of
the QE approximation is available when the fast reactions have no
joint reagents. In general case, we need either a numerical solver
for (\ref{EQforQE}) or some additional hypotheses about smallness.
Michaelis and Menten used, in addition to the QE approach, the
hypothesis about smallness of the amount of intermediate complexes.
This is the typical QSS hypothesis. The QE approximation was
modified and further developed by many authors. In particular, a
computational optimization approach for the numerical approximation
of slow attracting manifolds based on entropy-related and geometric
extremum principles for reaction trajectories was developed
\cite{Lebiedz2010}.

Of course, validity of all the simplification hypotheses is a
crucial question. For example, for the CO oxidation, if we accept
the hypothesis about the quasiequilibrium adsorption then we get a
simple dynamics which monotonically tends to the steady state. The
state of the surface is unambiguously presented as a continuous
function of the gas composition. The pure QSS hypothesis results for
the Langmuir-Hinshelwood reaction mechanism
(\ref{Langmuir-Hinshelwood}) without quasiequilibrium adsorption in
bifurcations and the multiplicity of steady states
\cite{Yablonskii1991}. The problem of validity of simplifications
cannot be solved as a purely theoretical question without the
knowledge of kinetic constants or some additional experimental~data.

\section{General Kinetics with Fast Intermediates
 Present in Small Amount \label{GeneralKinetics}}

\subsection{Stoichiometry of Complexes\label{CoplexStoi}}

In this Section, we return to the very general reaction network.

Let us call all the formal sums that participate in the
stoichiometric equations (\ref{stoichiometricequation}), the
{\em complexes}. The set of complexes for a given reaction
mechanism (\ref{stoichiometricequation}) is $\Theta_1, \ldots,
\Theta_q$. The number of complexes $q\leq 2m$ (two complexes
per elementary reaction, as the maximum) and it is possible
that $q < 2m$ because some complexes may coincide for different
reactions.

A complex $\Theta_i$ is a formal sum $\Theta_i=\sum_{j=1}^n \nu_{ij}
A_j= (\nu_i,A)$, where $\nu_i$ is a vector with coordinates
$\nu_{ij}$.

Each elementary reaction (\ref{stoichiometricequation}) may be
represented in the form $\Theta^-_{\rho}\to \Theta^+_{\rho}$, where
$\Theta^{\pm}_{\rho}$ are the complexes which correspond to the
right and the left sides (\ref{stoichiometricequation}). The whole
mechanism is naturally represented as a digraph of transformation of
complexes: Vertices are complexes and edges are reactions. This
graph gives a convenient tool for the reaction representation and is
often called the ``reaction graph".

Let us consider a simple example: 18 elementary reactions (9 pairs
of mutually reverse reactions) from the hydrogen combustion
mechanism (see, for example, \cite{Conaireatal2004}).
\begin{equation}\label{hyroexample}
\begin{array}{ll}
 {\rm H + O_2 \rightleftharpoons O + OH; \; }& {\rm O + H_2
\rightleftharpoons H + OH; }\\
  {\rm  OH + H_2 \rightleftharpoons H + H_2O;} \;&{\rm   O + H_2O \rightleftharpoons 2OH;}\\
 {\rm  HO_2 + H \rightleftharpoons H_2 + O_2;} \;&{\rm  HO_2 + H \rightleftharpoons 2OH;}\\
 {\rm  H + OH +M \rightleftharpoons H_2O +M;} \;& {\rm  H + O_2 +M \rightleftharpoons HO_2 +M; }\\
  {\rm H_2O_2 + H \rightleftharpoons H_2 + HO_2}&
\end{array}
\end{equation}
There are 16 different complexes here:
\begin{eqnarray*}&&{\rm \Theta_1=H + O_2,
\, \Theta_2= O + OH, \, \Theta_3=O + H_2, \, \Theta_4=H+OH,}\\
&&{\rm \Theta_5 =OH + H_2,
\rm \Theta_6=H + H_2O, \, \Theta_7= O + H_2O, \, \Theta_8=2OH, \,}\\
&&{\rm \Theta_9=HO_2 + H,
 \rm \Theta_{10}=H_2 + O_2, \, \Theta_{11}=H + OH +M, \,}\\
&&{\rm \Theta_{12}=H_2O +M, \, \rm \Theta_{13}=H + O_2 +M,\, \Theta_{14}=HO_2 +M,  }\\
 &&{\rm  \Theta_{15}=H_2O_2 + H, \,  \Theta_{16}=H_2 + HO_2} \,
\end{eqnarray*}
The reaction set (\ref{hyroexample}) can be represented as
\begin{eqnarray*}&& \Theta_1 \rightleftharpoons \Theta_2, \,  \Theta_3 \rightleftharpoons \Theta_4, \, \Theta_5\rightleftharpoons
\Theta_6, \, \Theta_7\rightleftharpoons \Theta_8
\rightleftharpoons \Theta_9 \leftrightharpoons \Theta_{10},  \\
&&  \Theta_{11}\rightleftharpoons \Theta_{12},
\,\Theta_{13}\rightleftharpoons \Theta_{14}, \,
\Theta_{15}\rightleftharpoons \Theta_{16}\,
\end{eqnarray*}
We can see that this digraph of transformation of complexes has
a very simple structure: There are five isolated pairs of
complexes and one connected group of four complexes.

\subsection{Stoichiometry of Compounds}

For each complex $\Theta_j$ we introduce an additional component
$B_j$, an intermediate compound and $B^{\pm}_{\rho}$ are those
compounds $B_j$ ($1 \leq j \leq q$), which correspond to the right
and left sides of reaction (\ref{stoichiometricequation}).

We call these components ``compounds'' following the English
translation of the original Michaelis-Menten paper
\cite{MichaelisMenten1913} and keep ``complexes'' for the formal
linear combinations $\Theta_j$.

An extended reaction mechanism includes two types of reactions:
Equilibration between a complex and its compound ($q$ reactions, one
for each complex)
\begin{equation}\label{equilibrationreact}
\Theta_j \rightleftharpoons B_j
\end{equation}
and transformation of compounds $B_{\rho}^- \to B_{\rho}^+$ ($m$
reactions, one for each elementary reaction from
(\ref{stoichiometricequation}). So, instead of the reaction
(\ref{stoichiometricequation}) we can write
\begin{equation}\label{stoichiometricequationcompaund}
\sum_i\alpha_{\rho i}A_i \rightleftharpoons B_{\rho}^- \to
B_{\rho}^+ \rightleftharpoons \sum_i \beta_{\rho i} A_i \,
\end{equation}
Further on we assume that if the input or output complexes coincide for two reactions
then the corresponding equilibration reactions also coincide. This hypothesis ``one
complex --- one compound'' is convenient for the following formalism. Indeed, let us
assume that there are If we assume that the reactions (\ref{equilibrationreact}) are two
equilibration reactions for a complex $\Theta$: $\Theta \rightleftharpoons B_1$ and
$\Theta \rightleftharpoons B_2$. If these reactions are in equilibrium then the
concentrations of $B_1$ and $B_2$ are proportional (see the free energy (\ref{FreeEn1})
and the equilibria description (\ref{equilibrationEq}) below). Of course, it is not
forbidden to introduce several compounds for one complex and if it is necessary then the
corresponding modification of the formalism is quite simple.

It is useful to visualize the reaction scheme. In
Figure~\ref{n-tail} we represent the $2n$-tail scheme of an
elementary reaction sequence (\ref{stoichiometricequationcompaund})
which is an extension of the elementary reaction
(\ref{stoichiometricequation}).

\begin{figure}[H]
\centering{
\includegraphics[width=0.5\textwidth]{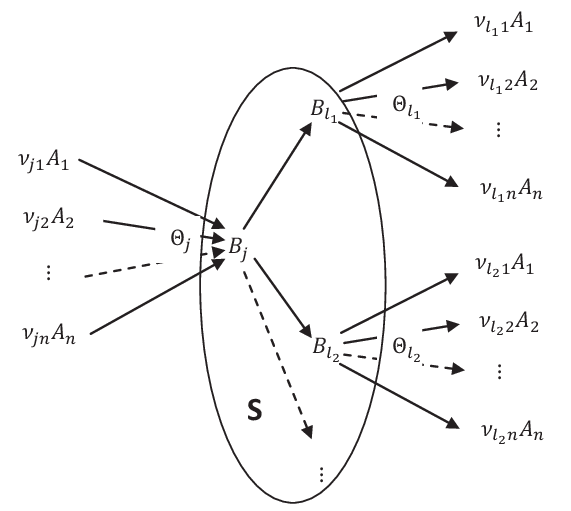}
}\caption{A multichannel view on the complex transformation. The
hidden reactions between compounds are included in an oval
$\mathbf{S}$. \label{multichannel}}
\end{figure}

The reactions between compounds may have several channels
(Figure~\ref{multichannel}): One complex may transform to several
other complexes.

The reaction mechanism is a set of multichannel transformations
(Figure~\ref{multichannel}) for all input complexes. In
Figure~\ref{multichannel} we grouped together the reactions with the
same input complex. Another representation of the reaction mechanism
is based on the grouping of reactions with the same output complex.
Below, in the description of the complex balance condition, we use
both representations.

The extended list of components includes $n+q$ components: $n$
initial species $A_i$ and $q$ compounds $B_j$. The corresponding
composition vector $N^{\oplus}$ is a direct sum of two vectors, the
composition vector for initial species, $N$, with coordinates $N_i$
($i=1, \ldots , n$) and the composition vector for compounds,
$\Upsilon$, with coordinates $\Upsilon_j$ ($j=1, \ldots , q$):
$N^{\oplus}=N\oplus \Upsilon$.

The space of composition vectors $E$ is a direct sum of
$n$-dimensional $E_A$ and $q$-dimensional $E_B$: $E=E_A\oplus E_B$.

For concentrations of $A_i$ we use the notation $c_i$ and for
concentrations of $B_j$ we use $\varsigma_j$.

The stoichiometric vectors for reactions $\Theta_j
\rightleftharpoons B_j$ (\ref{equilibrationreact}) are direct sums:
$g^j=-\nu_j \oplus e_j$, where $e_j$ is the $j$th standard basis
vector of the space $R^q=E_B$, the coordinates of $e_j$ are
$e_{jl}=\delta_{jl}$:
\begin{equation}\label{stoichoiomVectEquilib}
g^j=(-\nu_{j1},-\nu_{j2}, \ldots ,-\nu_{jn},\underbrace{0,\ldots ,
0, 1}_l,0,\ldots ,0)
\end{equation}

The stoichiometric vectors of equilibration reactions
(\ref{equilibrationreact}) are linearly independent because there
exists exactly one vector for each $l$.

The stoichiometric vectors $\gamma^{jl}$ of reactions $B_j \to B_l$
belong entirely to $E_B$. They have $j$th coordinate $-1$, $l$th
coordinate $+1$ and other coordinates are zeros.

To exclude some degenerated cases a hypothesis of {\em weak
reversibility} is accepted. Let us consider a digraph with vertices
$\Theta_i$ and edges, which correspond to reactions from
(\ref{stoichiometricequation}). The system is weakly reversible if
for any two vertices $\Theta_i$ and $\Theta_j$, the existence of an
oriented path from $\Theta_i$ to $\Theta_j$ implies the existence of
an oriented path from $\Theta_j$ to $\Theta_i$.

Of course, this weak reversibility property is equivalent to weak
reversibility of the reaction network between compounds $B_j$.

\subsection{Energy, Entropy and Equilibria of Compounds}

In this section, we define the free energy of the system. The basic
hypothesis is that the compounds are the small admixtures to the
system, that is, the amount of compounds $B_j$ is much smaller than
amount of initial components $A_i$. Following this hypothesis, we
neglect the energy of interaction between compounds, which is
quadratic in their concentrations because in the low density limit
we can neglect the correlations between particles if the potential
of their interactions decay sufficiently fast when the distance
between particles goes to $\infty$ \cite{Balescu}. We take the
energy of their interaction with $A_i$ in the linear approximation,
and use the perfect entropy for $B_i$. These standard assumptions
for a small admixtures give for the free energy:
\begin{equation}\label{FreeEn1}
F=Vf(c,T)+VRT \sum_{j=1}^q \varsigma_j \left(\frac{u_j(c,T)}{RT}+\ln
\varsigma_j-1\right)
\end{equation}

The thermodynamic equilibrium of a system of reactions is the free energy minimizer under
given values of the stoichiometric conservation laws. If $L$ is the linear span of the
stoichiometric vectors then $L^{\perp}$ is the space of stoichiometric conservation laws:
$(b,N)$ is a linear conservation law for any $b\in L^{\perp}$. The thermodynamic
equilibria of a system of reactions form a manifold in the cone of positive
concentrations. This manifold is parameterized by the values of linear conservation laws
from a basis of the space $L^{\perp}$. If a reaction mechanism is split into several
reaction subsystems then the manifold of equilibria is the intersection of the manifolds
of the equilibria for subsystems. Such a representation may be convenient when the
manifolds of equilibria for subsystems can be found explicitly, whereas their
intersection has not so simple representation. Let us describe two equilibrium manifolds,
(i) for the system of equilibration reactions (\ref{equilibrationreact}) and (ii) for the
system of transitions between compounds, $B_j\to B_l$.  The assumption of weak
reversibility of the transitions between compounds will be necessary to provide
equivalence between the thermodynamic and kinetic equilibria because it is necessary for
existence of positive kinetic equilibria of first order kinetics.

Let us introduce the {\em standard equilibrium} concentrations for
$B_j$. Due to the Boltzmann distribution ($\exp(-u/RT)$) and formula
(\ref{FreeEn1})
\begin{equation}\label{StandEquili}
\varsigma_j^*(c,T)=\frac{1}{Z}\exp\left(-\frac{u_j(c,T)}{RT}\right)
\end{equation}
where $1/Z$ is the normalization factor. Let us select here the
normalization $Z=1$ and write:
\begin{equation}\label{FreeEn2}
F=Vf(c,T)+VRT \sum_{j=1}^q \varsigma_j \left(\ln
\left(\frac{\varsigma_j}{\varsigma_j^*(c,T)}\right)-1\right) \
\end{equation}
We assume that the standard equilibrium concentrations
$\varsigma_j^*(c,T)$ are much smaller than the concentrations of
$A_i$. It is always possible because functions $u_j$ are defined up
to an additive constant.

The formula for free energy is necessary to define the fast
equilibria (\ref{equilibrationreact}). Such an equilibrium is the
minimizer of the free energy on the straight line parameterized by
$a$: $c_i=c_i^0-a \nu_{ji}$, $\varsigma_j=a$.

If we neglect the products $\varsigma_j \partial
\varsigma_j^*(c,T)/\partial c_i$ as the second order small
quantities then the minimizers have the very simple form:
\begin{equation}\label{equilibrationEqMUMU}
\vartheta_j=\sum_i \nu_{ji} \frac{\mu_i(c,T)}{RT}
\end{equation}
or
\begin{equation}\label{equilibrationEq}
\varsigma_j=\varsigma^*_j(c,T)\exp\left(\frac{\sum_i \nu_{ji}
\mu_i(c,T)}{RT}\right)
\end{equation}
where $$\mu_i=\frac{\partial f(c,T)}{\partial c_i}$$ is the chemical
potential of $A_i$ and
$$\vartheta_j=\ln\left(\frac{\varsigma_j}{\varsigma^*_j}\right)$$
($RT\vartheta_j=\frac{1}{V}\frac{\partial F}{\partial \varsigma_j}$
is the chemical potential of $B_j$).

Equation (\ref{equilibrationEq}) represents the  equilibria of the system of reactions
equilibria (\ref{equilibrationreact}) parameterized by $c$ (i.e. by the concentrations of
the components $A_i$).

The thermodynamic equilibrium of the system of reactions $B_j \to
B_l$ that corresponds to the
reactions~(\ref{stoichiometricequationcompaund}) is the free energy
minimizer under given values of the conservation laws.

For the systems with fixed volume, the {\em stoichiometric conservation laws} of the
monomolecular system of reaction are sums of the concentrations of $B_j$ which belong to
the connected components of the reaction graph. Under the hypothesis of weak
reversibility there is no other linear conservation law. Let the graph of reactions $B_j
\to B_l$ have $d$ connected components $C_s$ and let $V_s$ be the set of indexes of those
$B_j$ which belong to $C_s$: $B_j \in C_s$ if and only if $j \in V_s$. For each $C_s$
there exists a stoichiometric conservation~law
\begin{equation}\label{LinearConservation}
\beta_s=\sum_{j\in V_s}\varsigma_j=const
\end{equation}

For any set of positive values of $\beta_s$ ($s=1, \ldots , d$) and given $c,T$ there
exists a unique conditional minimizer $\varsigma^{\rm eq}_j$ of the free energy
(\ref{FreeEn2}): For the compound $B_j$ from the $s$th connected component ($j \in V_s$)
this equilibrium concentration is
\begin{equation}\label{compoundEq}
\varsigma^{\rm eq}_j=\beta_s\frac{\varsigma_j^*(c,T)}{\sum_{l\in V_s
}\varsigma_j^*(c,T)}
\end{equation}

The positive values of concentrations $\varsigma_j$ are the
equilibrium concentrations (\ref{compoundEq}) for some values of
$\beta_s$ if and only if for any $s=1, \ldots , d$ and all $j,l\in
V_s$
\begin{equation}\label{equilibriumcompoundTheta}
\vartheta_j=\vartheta_l
\end{equation}
($\vartheta_j=\ln(\varsigma_j/\varsigma_j^*)$). This means that
compounds are in equilibrium in every connected component $C_s$ the
chemical potentials of compounds coincide in each component $C_s$.
The system of equations (\ref{equilibriumcompoundTheta}) together
with the equilibrium conditions (\ref{equilibrationEq}) constitute
the equilibrium of the systems. All the equilibria form a linear
subspace in the space with coordinates $\mu_i/RT$ ($i=1, \ldots, n$)
and $\vartheta_j$ ($j=1, \ldots , q$).

In the expression for the free energy (\ref{FreeEn2}) we do not
assume anything special about free energy of the mixture of $A_i$.
The density of this free energy, $f(c,T)$, may be an arbitrary
smooth function (later, we will add the standard assumption about
convexity of $f(c,T)$ as a function of $c$). For the compounds
$B_i$, we assume that they form a very small addition to the mixture
of $A_i$, neglect all quadratic terms in concentrations of $B_i$ and
use the entropy of the perfect systems, $p\ln p$, for this small
admixture.

This approach results in the explicit expressions for the fast
equilibria (\ref{equilibrationEq}) and expression of the equilibrium
compound concentrations through the values of the stoichiometric
conservation laws (\ref{compoundEq}).

\subsection{Markov Kinetics of Compounds}

For the kinetics of compounds transformations $B_j \to B_l$, the
same hypothesis of the smallness of concentrations leads to the only
reasonable assumption: The linear (monomolecular) kinetics with the
rate constant $\kappa_{lj} >0$. This ``constant'' is a function of
$c,T$: $\kappa_{lj}(c,T)$. The order of  indexes at $\kappa$ is
inverse to the order of them in reaction: $\kappa_{lj}=\kappa_{l
\leftarrow j}$.

The master equation for the concentration of $B_j$ gives:
\begin{equation}\label{MasterEq}
\frac{\D \varsigma_j}{\D t}=\sum_{l, \, l\neq j} \left(\kappa_{jl}
\varsigma_l-\kappa_{lj}\varsigma_j\right)
\end{equation}
It is necessary to find when this kinetics respect thermodynamics,
{\em i.e.},  when the free energy decreases due to the system
(\ref{MasterEq}). The necessary and sufficient condition for
matching the kinetics and thermodynamics is: The standard
equilibrium $\varsigma^*$ (\ref{StandEquili}) should be an
equilibrium for (\ref{MasterEq}), that is, for every $j=1, \ldots ,
q$
\begin{equation}\label{MarkovEqEquilibriumCond}
\sum_{l, \, l\neq j} \kappa_{jl} \varsigma_l^*=\sum_{l, \, l\neq j}
\kappa_{lj}\varsigma_j^*
\end{equation}
This condition is necessary because the standard equilibrium is the
free energy minimizer for given $c,T$ and $\sum_j\varsigma_j =
\sum_j \varsigma_j^*$. The sum $\sum_j\varsigma_j$ conserves due to
(\ref{MasterEq}). Therefore, if we assume that $F$ decreases
monotonically due to (\ref{MasterEq}) then  the point of conditional
minimum of $F$ on the plane $\sum_j\varsigma_j =const$ (under given
$c,T$) should be an equilibrium point for this kinetic system. This
condition is sufficient due to the Morimoto $H$-theorem (see
Appendix 2).

For a weakly reversible system, the set of the conditional
minimizers of the free energy (\ref{compoundEq}) coincides with with
the set of positive equilibria for the master equations
(\ref{MasterEq}) (see Equation (\ref{compoundEqA}) in Appendix 2).

\subsection{Thermodynamics and Kinetics of the
Extended System}

In this section, we consider the complete extended system, which
consists of species $A_i$ ($i=1, \ldots , n$) and compounds $B_j$
($j=1,\ldots , q$) and includes reaction of equilibration
(\ref{equilibrationreact}) and transformations of compounds $B_j \to
B_l$ which correspond to the reactions
(\ref{stoichiometricequationcompaund}).

Thermodynamic properties of the system are summarized in the free
energy function (\ref{FreeEn2}). For kinetics of compounds we accept
the Markov model (\ref{MasterEq}) with the equilibrium condition
(\ref{MarkovEqEquilibriumCond}), which guarantees matching between
thermodynamics and kinetics.

For the equilibration reactions (\ref{equilibrationreact}) we select
a very general form of the kinetic law. The only requirement is:
This reaction should go to its equilibrium, which is described as
the conditional minimizer of free energy $F$
(\ref{equilibrationEq}). For each reaction
$\Theta_j\rightleftharpoons B_j$ (where the complex is a formal
combination: $\Theta_j=\sum_i \nu_{ji} A_i$)  we introduce the
reaction rate $w_j$. This rate should be positive if
\begin{equation}\label{equilibrationLess}
\vartheta_j< \sum_i \nu_{ji} \frac{\mu_i(c,T)}{RT}
\end{equation}
and negative if
\begin{equation}\label{equilibrationGross}
\vartheta_j> \sum_i \nu_{ji} \frac{\mu_i(c,T)}{RT}
\end{equation}

The general way to satisfy these requirement is to select $q$
continuous function of real variable $w_j(x)$, which are negative if
$x>0$ and positive if $x<0$. For the equilibration rates we take
\begin{equation}\label{equilibrationRates}
w_j=w_j\left(\vartheta_j - \sum_i \nu_{ji}
\frac{\mu_i(c,T)}{RT}\right)
\end{equation}

If several dynamical systems defined by equations $\dot{x}=J_1$, ...
$\dot{x}=J_v$ on the same space have the same Lyapunov function $F$,
then for any conic combination $J=\sum_k a_k J_k$ ($a_k\geq 0$,
$\sum_k a_k >0$) the dynamical system $\dot{x}=J$ also has the
Lyapunov function $F$.

The free energy (\ref{FreeEn2}) decreases monotonically due to any
reaction $\Theta_j\rightleftharpoons B_j$ with reaction rate $w_j$
(\ref{equilibrationRates}) and also due to the Markov kinetics
(\ref{MasterEq}) with the equilibrium condition
(\ref{MarkovEqEquilibriumCond}). Therefore, the free energy
decreases monotonically due to the following kinetic system:
\begin{equation}\label{GrossEquationSys}
\begin{split}
&\frac{\D c_i}{\D t}= - \sum_{j=1}^q \nu_{ji} w_j \,  \\
&\frac{\D \varsigma_j}{\D t}= w_j + \sum_{l, \, l\neq j}
\left(\kappa_{jl} \varsigma_l-\kappa_{lj}\varsigma_j\right)
\end{split}
\end{equation}
where the coefficients $\kappa_{jl}$ satisfy the matching condition
(\ref{MarkovEqEquilibriumCond}).

This general system (\ref{GrossEquationSys}) describes kinetics of
extended system and satisfies all the basic conditions
(thermodynamics and smallness of compound concentrations). In the
next sections we will study the QE approximations to this system and
exclude the unknown functions $w_j$ from it.

\subsection{QE Elimination of Compounds and the Complex Balance Condition \label{Sec:QEelimComp}}

In this section, we use the QE formalism developed for chemical
kinetics in Section~\ref{ChemKinQEApprox} for simplification of the
compound kinetics.

First of all, let us describe $L^{\perp}$, where the space $L$ is
the subspace in the extended concentration space spanned by the
stoichiometric vectors of fast equilibration reactions
(\ref{equilibrationreact}). The stoichiometric vector for the
equilibration reactions have a very special structure
(\ref{stoichoiomVectEquilib}). Dimension of the space $L$ is equal
to the number of complexes: $ \dim L=q$. Therefore, dimension of
$L^{\perp}$ is equal to the number of components $A_i$: $\dim
L^{\perp}=n$. For each $A_i$ we will find a vector $b_i\in
L^{\perp}$ that has the following first $n$ coordinates:
$b_{ik}=\delta_{ik}$ for $k=1,\ldots, n$. The condition
$(b_i,g_j)=0$ gives immediately: $b_{i,n+j}=\nu_{ji}$. Finally,
\begin{equation}\label{basisLperpcomlex}
b_i=(\overbrace{\underbrace{0,\ldots
,0,1}_i,0,\ldots,0}^n,\nu_{1i},\nu_{2i},\ldots, \nu_{qi})
\end{equation}
The corresponding slow variables are
\begin{equation}\label{slowvariablescomplexQE}
b_i(c,\varsigma)=c_i+\sum_j \varsigma_j \nu_{ji}
\end{equation}

In the QE approximation all $w_j=0$ and the kinetic equations
(\ref{GrossEquationSys}) give in this approximation
\begin{equation}\label{QEequationkinSlowComp}
\frac{\D b_i}{\D t}= \sum_{lj, \,l\neq  j}( \kappa_{jl}
\varsigma_l-\kappa_{lj} \varsigma_j )\nu_{ji}
\end{equation}
In these equations, we have to use the dependence $\varsigma(b)$.
Here we use the QSS Michaelis and Menten assumption: The compounds
are present in small amounts
$$c_i
\gg \varsigma_j$$ In this case, we can take $b_i$ instead of $c_i$
({\em i.e.},  take $\mu(b,T)$ instead of $\mu(c,T)$) in the formulas
for equilibria~(\ref{equilibrationEq}):
\begin{equation}\label{equilibrationEqSLOW}
\varsigma_j=\varsigma^*_j(b,T)\exp\left(\frac{\sum_i \nu_{ji}
\mu_i(b,T)}{RT}\right)
\end{equation}

In the final form of the QE kinetic equation there remain two
``offprints'' of the compound kinetics: Two sets of functions
$\varsigma^*_j(b,T)\geq 0$ and $\kappa_{jl}(b,T)\geq 0$. These
functions are connected by the identity
(\ref{MarkovEqEquilibriumCond}). The final form of the equations is
\begin{equation}\label{QEequationkinSlowCompFinal}
\begin{split}
\frac{\D b_i}{\D t}=& \sum_{lj, \,l\neq  j}\left( \kappa_{jl}
\varsigma^*_l(b,T)\exp\left(\frac{\sum_i \nu_{li}
\mu_i(b,T)}{RT}\right)-\kappa_{lj}
\varsigma^*_j(b,T)\exp\left(\frac{\sum_i \nu_{ji}
\mu_i(b,T)}{RT}\right) \right)\nu_{ji}
\end{split}
\end{equation}
The identity (\ref{MarkovEqEquilibriumCond}), $\sum_{l, \, l\neq j}
\kappa_{jl} \varsigma_l^*=\sum_{l, \, l\neq j}
\kappa_{lj}\varsigma_j^*$, provides a sufficient condition for
decreasing of free energy due to the kinetic equations
(\ref{QEequationkinSlowCompFinal}). This is a direct consequence of
two theorem: The theorem about the preservation of entropy
production in the QE approximations (see Section~\ref{General} and
Appendix 1) and the Morimoto $H$-theorem (see Appendix 2). Indeed,
in the QE state the equilibrated reactions
(\ref{equilibrationreact}) $\Theta_j \rightleftharpoons B_j$ do not
produce entropy and all changes in the total free energy are caused
by the Markov kinetics $B_i \to B_j$. Due to the Morimoto
$H$-theorem this change is negative: The Markov kinetics decrease
the perfect free energy of compounds and do not affect the free
energy of $A_i$. In the QE approximation, the concentrations of
$A_i$ are changing together with concentrations of $B_j$ because of
the equilibrium conditions for reactions $\Theta_j
\rightleftharpoons B_j$. Due to the theorem of preservation of the
entropy production, the time derivative of the total free energy in
this QE dynamics coincides with the time derivative of the free
energy of $B_j$ due to Markov kinetics. In addition to this proof,
in Section~\ref{Sec:GenKinThermod} below we give the explicit
formula for entropy production in (\ref{QEequationkinSlowCompFinal})
and direct proof of its positivity.

Let us stress that the functions $\varsigma^*_j(b,T)$ and
$\kappa_{jl}(b,T)$ participate in equations
(\ref{QEequationkinSlowCompFinal}) and in identity
(\ref{MarkovEqEquilibriumCond}) in the form of the product. Below we
use for this product a special notation:
\begin{equation}\label{kinetic multiplier}
\varphi_{jl}(b,T)=\kappa_{jl}(b,T)\varsigma^*_l(b,T) \; (j\neq l)
\end{equation}
We call this function $\varphi_{jl}(b,T)$ the {\em kinetic factor}.
The identity (\ref{MarkovEqEquilibriumCond}) for the kinetic factor
is
\begin{equation}\label{kinetic multiplier complex balance}
\sum_{l, \, l\neq j} \varphi_{jl}(b,T)= \sum_{l, \, l\neq j}
\varphi_{lj}(b,T)  \mbox{ for all }  j
\end{equation}

We call the {\em thermodynamic factor} (or the Boltzmann factor) the
second multiplier in the reaction~rates
\begin{equation}\label{BoltzmannFactor}
\Omega_l(b,T)=\exp\left(\frac{\sum_i \nu_{li} \mu_i(b,T)}{RT}\right)
\end{equation}
In this notation, the kinetic equations
(\ref{QEequationkinSlowCompFinal}) have a simple form
\begin{equation}\label{QEequationkinSlowCompFinalSimple}
\frac{\D b_i}{\D t}= \sum_{lj, \,l\neq  j}( \varphi_{jl}(b,T)
\Omega_l(b,T)-\varphi_{lj}(b,T) \Omega_j(b,T) )\nu_{ji}
\end{equation}

The general equations (\ref{QEequationkinSlowCompFinalSimple}) have
the form of ``sum over complexes''. Let us return to the more usual
``sum over reactions'' form. An elementary reaction corresponds to
the pair of complexes $\Theta_l,\Theta_j$
(\ref{stoichiometricequationcompaund}). It has the form $\Theta_l
\to \Theta_j$ and the reaction rate is $r=\varphi_{jl}\Omega_l$. In
the right hand side in (\ref{QEequationkinSlowCompFinalSimple}) this
reaction appears twice: first time with sign ``$+$'' and the vector
coefficient $\nu_{j}$ and the second time with sign ``$-$'' and the
vector coefficient $\nu_{l}$. The stoichiometric vector of this
reaction is $\gamma=\nu_{j}-\nu_{l}$. Let us enumerate the
elementary reactions by index $\rho$, which corresponds to the pair
$(j,l)$. Finally, we transform
(\ref{stoichiometricequationcompaund}) into the sum over reactions
form
\begin{equation}\label{QEequationkinSlowCompFinalSimpleEQ}
\begin{split}
\frac{\D b_i}{\D t}&= \sum_{l,j, \,l\neq  j}\varphi_{jl}(b,T)
\Omega_l(b,T)(\nu_{ji}-\nu_{li}) \\
&=\sum_{\rho}\varphi_{\rho}(b,T) \Omega_{\rho}(b,T)\gamma_{{\rho}i}
\end{split}
\end{equation}
In the vector form it looks as follows:
\begin{equation}\label{QEequationkinSlowCompFinalSimpleEQvec}
\frac{\D b}{\D t}= \sum_{\rho}\varphi_{\rho}(b,T)
\Omega_{\rho}(b,T)\gamma_{{\rho}}
\end{equation}

\subsection{The Big Michaelis-Menten-Stueckelberg Theorem \label{Sec:TheTeorem}}

Let us summarize the results of our analysis in one statement.

Let us consider the reaction mechanism illustrated by
Figure~\ref{multichannel} (\ref{stoichiometricequationcompaund}):
\begin{equation*}
\sum_i\alpha_{\rho i}A_i \rightleftharpoons B_{\rho}^- \to
B_{\rho}^+ \rightleftharpoons \sum_i \beta_{\rho i} A_i
\end{equation*}
under the following asymptotic assumptions:
\begin{enumerate}
\item{Concentrations of the compounds $B_{\rho}$ are close
    to their quasiequilibrium values
    (\ref{equilibrationEqSLOW})
\begin{equation*}
\varsigma_j=(1+\delta)\varsigma_j^{\rm
QE}=(1+\delta)\varsigma^*_j(b,T)\exp\left(\frac{\sum_i \nu_{ji}
\mu_i(b,T)}{RT}\right)\, , \;\; \delta \ll 1
\end{equation*}
(this may be due to the fast reversible reactions in
(\ref{stoichiometricequationcompaund}));}\vspace{-10pt}
\item{Concentrations of the compounds $B_{\rho}$ are much
    smaller than the concentrations of the components
    $A_i$: There is a small positive parameter
    $\varepsilon\ll 1$, $\varsigma^*_j=\varepsilon \xi^*_j$
    and $\xi^*_j$ do not depend on $\varepsilon$;}\vspace{-10pt}
\item{Kinetics of transitions between compounds $B_i\to
    B_j$ is linear (Markov) kinetics with reaction rate
    constants $k_{ji}=\frac{1}{\varepsilon}\kappa_{ji}$.}
\end{enumerate}

Under these assumptions, in the asymptotic $\delta, \varepsilon \to
0$, $\delta, \varepsilon > 0$ kinetics of components $A_i$ may be
described by the reaction mechanism
\begin{equation*}
\sum_i\alpha_{\rho i}A_i  \to  \sum_i \beta_{\rho i} A_i
\end{equation*}
with the reaction rates
\begin{equation*}
r_{\rho}=\varphi_{\rho}\exp\left(\frac{(\alpha_{\rho},
{\mu})}{RT}\right)
\end{equation*}
where the kinetic factors $\varphi_{\rho}$ satisfy the  condition
(\ref{kinetic multiplier complex balance}):
\begin{equation*}
\sum_{\rho, \,\alpha_{\rho}=\mathbf{v}} \varphi_{\rho}\equiv
\sum_{\rho, \,\beta_{\rho}=\mathbf{v}} \varphi_{\rho}
\end{equation*}
for any vector $\mathbf{v}$ from the set of all vectors
$\{\alpha_{\rho}, \beta_{\rho}\}$. This statement includes the
generalized mass action law for $r_{\rho}$ and the balance identity
for kinetic factors that is sufficient for the entropy growth as it
is shown in the next Section~\ref{Sec:GenKinThermod}.

\section{General Kinetics and Thermodynamics \label{Sec:GenKinThermod}}

\subsection{General Formalism}

To produce the general kinetic law and the complex balance
conditions, we use ``construction staging'': The intermediate
complexes with fast equilibria, the Markov kinetics and other
important and interesting physical and chemical hypothesis.

In this section, we delete these construction staging and start from
the forms (\ref{BoltzmannFactor}),
(\ref{QEequationkinSlowCompFinalSimpleEQvec}) as the basic laws. We
use also the complex balance conditions (\ref{kinetic multiplier
complex balance}) as a hint for the general conditions which
guarantee accordance between kinetics and thermodynamics.

Let us consider a domain $U$ in $n$-dimensional real vector space
$E$ with coordinates $N_1,\ldots, N_n$. For each $N_i$ a symbol
(component) $A_i$ is given. A dimensionless entropy (or {\em free
entropy}, for example, Massieu, Planck, or Massieu-Planck potential
which correspond to the selected conditions \cite{Callen1985})
$S(N)$ is defined in $U$. ``Dimensionless'' means that we use $S/R$
instead of physical $S$. This choice of units corresponds to the
informational entropy ($p\ln p$ instead of $k_{\rm B} p\ln p$).

The dual variables, potentials, are defined as the partial
derivatives of $S$:
\begin{equation}
\check{\mu}_i=-\frac{\partial S}{\partial N_i}
\end{equation}
{\bf Warning}: This definition differs from the chemical potentials
(\ref{chemical potential}) by the factor ${1}/{RT}$: For constant
volume the Massieu-Planck potential is $-F/T$ and we, in addition,
divide it on $R$. On the other hand, we keep the same sign as for
the chemical potentials, and this differs from the standard Legendre
transform for $S$. (It is the Legendre transform for function $-S$).

The reaction mechanism is defined by the stoichiometric equations
(\ref{stoichiometricequation})
$$\sum_i\alpha_{\rho i}A_i \to \sum_i \beta_{\rho i} A_i $$
($\rho =1, \ldots, m$). In general, there is no need to assume that
the stoichiometric coefficients $\alpha_{\rho i},\beta_{\rho i}$
are~integers.

The assumption that they are nonnegative, $\alpha_{\rho i}\geq
0,\beta_{\rho i} \geq 0$, may be needed to prove that the kinetic
equations preserve positivity of $N_i$. If $N_i$ is the number of
particles then it is a natural assumption but we can use other
extensive variables instead, for example, we included energy in the
list of variables to describe the non-isothermal processes
\cite{BykGOrYab1982}. In this case, the coefficient $\alpha_U$ for
the energy component $A_U$ in an exothermic reaction is negative.

So, for variables that are positive (bounded from below) by their
physical sense, we will use the inequalities $\alpha_{\rho i}\geq
0,\beta_{\rho i} \geq 0$, when necessary, but in general, for
arbitrary extensive variables, we do not assume positivity of
stoichiometric coefficients. As it is usually, the stoichiometric
vector of reaction is  $\gamma_{\rho}=\beta_{\rho}-\alpha_{\rho}$
(the ``gain minus loss'' vector).

For each reaction, a {\em nonnegative} quantity, reaction rate
$r_{\rho}$ is defined. We assume that this quantity has the
following structure:
\begin{equation}\label{GeneralReactionRate}
r_{\rho}=\varphi_{\rho}\exp(\alpha_{\rho}, \check{\mu} )
\end{equation}
where $(\alpha_{\rho}, \check{\mu} )=\sum_i \alpha_{\rho i}
\check{\mu}_i$.

In the standard formalism of chemical kinetics the reaction rates
are intensive variables and in kinetic equations for $N$ an
additional factor---the volume---appears. For heterogeneous systems,
there may be several ``volumes'' (including interphase surfaces).

Each reaction has it own ``volume'', an extensive variable
$V_{\rho}$ (some of them usually coincide), and we can write

\begin{equation}\label{generalKINURpolyvolume}
\frac{\D N}{\D t}=\sum_{\rho}V_{\rho} \gamma_{\rho}
\varphi_{\rho}\exp(\alpha_{\rho}, \check{\mu} )
\end{equation}

In these notations, both the kinetic and the Boltzmann factors are
intensive (and local) characteristics of the system.

Let us, for simplicity of notations, consider a system with one
volume, $V$  and write
\begin{equation}\label{generalKINUR}
\frac{\D N}{\D t}=V \sum_{\rho}\gamma_{\rho}
\varphi_{\rho}\exp(\alpha_{\rho}, \check{\mu} )
\end{equation}

Below we use the form (\ref{generalKINUR}). All our results will
hold also for the multi-volume systems
(\ref{generalKINURpolyvolume}) under one important assumption: The
elementary reaction
$$\sum_i\alpha_{\rho i}A_i \to \sum_i \beta_{\rho i} A_i \, $$
goes in the same volume as the reverse reaction
$$\sum_i \beta_{\rho i}A_i \to \sum_i \alpha_{\rho i}  A_i \, $$
or symbolically
\begin{equation}\label{plusminusvolume}
V_{\rho}^+=V_{\rho}^-
\end{equation}
If this condition (\ref{plusminusvolume}) holds then the detailed
balance conditions and the complex balance conditions will hold
separately in all volumes $V_{\rho}$.

An important particular case of (\ref{generalKINUR}) gives us the
Mass Action Law. Let us take the perfect free entropy
\begin{equation}\label{PerfectFreeEntropy}
S=-\sum_iN_i\left(\ln\left(\frac{c_i}{c_i^*}\right)-1\right)
\end{equation}
where $c_i=N_i/V\geq 0$ are concentrations and $c_i^*>0$ are the
standard equilibrium concentrations. Under isochoric conditions,
$V=const$, there is no difference between the choice of the main
variables, $N$~or~$c$.

For the perfect function (\ref{PerfectFreeEntropy})
\begin{equation}\label{MALmu}
\check{\mu}_i=\ln\left(\frac{c_i}{c_i^*}\right) \, , \;
\exp(\alpha_{\rho}, \check{\mu} )=\prod_i
\left(\frac{c_i}{c_i^*}\right)^{\alpha_{\rho i}}
\end{equation}
and for the reaction rate function (\ref{GeneralReactionRate}) we
get
\begin{equation}\label{MALreaction rate}
r_{\rho}=\varphi_{\rho}\prod_i
\left(\frac{c_i}{c_i^*}\right)^{\alpha_{\rho i}}
\end{equation}
The standard assumption for the Mass Action Law in physics and
chemistry is that $\varphi$ and $c^*$ are functions of temperature:
$\varphi_{\rho}=\varphi_{\rho}(T)$ and $c^*_i=c^*_i(T)$. To return
to the kinetic constants notation (\ref{MAL}) we should~write:
$$\frac{\varphi_{\rho}}{\prod_i {c_i^*}^{\alpha_{\rho i}}}=k_{\rho}$$

Equation (\ref{generalKINUR}) is the general form of the kinetic
equation we would like to study. In many senses, this form is too
general before we impose restrictions on the values of the kinetic
factors. For physical and chemical systems, thermodynamics is a
source of restrictions:
\begin{enumerate}
\item{The energy of the Universe is constant.} \vspace{-10pt}
\item{The entropy of the
Universe tends to a maximum.}
\end{enumerate}\vspace{-10pt}
(R. Clausius, 1865 \cite{Clausius}.)

\vspace{10pt} The first sentence should be extended: The kinetic
equations should respect several conservation laws: Energy, amount
of atoms of each kind (if there is no nuclear reactions in the
system) conservation of total probability and, sometimes, some other
conservation laws. All of them have the form of conservation of
values of some linear functionals: $l(N)=const$. If the input and
output flows are added to the system~then $$\frac{\D l(N)}{\D
t}=Vv^{\rm in}l^{\rm in}-v^{\rm out}l(N)$$ where $v^{\rm in,out}$
are the input and output fluxes per unit volume, $l^{\rm in}$ are
the input densities (concentration). The standard requirement is
that every reaction respects all these conservation laws. The formal
expression of this requirement is:
\begin{equation}\label{stoichiometricConservation}
l(\gamma_{\rho})=0 \mbox{ for all } \rho
\end{equation}
There is a special term for this conservation laws: The {\em
stoichiometric conservation laws}. All the main conservation laws
are assumed to be the stoichiometric ones.

Analysis of the stoichiometric conservation laws is a simple linear
algebra task: We have to find the linear functionals that annulate
all the stoichiometric vectors $\gamma_{\rho}$. In contrast, entropy
is not a linear function of $N$ and analysis of entropy production
is not so simple.

In the next subsection we discuss various conditions which guarantee
the positivity of entropy production in kinetic equations
(\ref{generalKINUR}).

\subsection{Accordance Between Kinetics and Thermodynamics \label{Sec:Accordance}}

\subsubsection{6.2.1. General Entropy Production Formula}

Let us calculate $\D S/ \D t$ due to equations (\ref{generalKINUR}):
\begin{equation}\label{entropyproductionGNEKIN}
\begin{split}
\frac{\D S}{\D t}&=\sum_i \frac{\partial S}{\partial N_i} \frac{\D N_i}{\D t} \\
 &=- \sum_i \check{\mu}_i V \sum_{\rho}\gamma_{\rho i} \varphi_{\rho}\exp(\alpha_{\rho},
\check{\mu} ) \\
 &=-V \sum_{\rho }(\gamma_{\rho }, \check{\mu})\varphi_{\rho}\exp(
\alpha_{\rho }, \check{\mu})
\end{split}
\end{equation}

An auxiliary function $\theta(\lambda) $ of one variable $\lambda\in
[0,1]$ is convenient for analysis of $\D S/ \D t$ (it was studied by
Rozonoer and Orlov \cite{OrlovRozonoer1984}, see also \cite{G1}:
\begin{equation}\label{auxtheta}
\theta(\lambda)=\sum_{\rho}\varphi_{\rho}\exp[(\check{\mu},(\lambda
\alpha_{\rho}+(1-\lambda)\beta_{\rho}))]
\end{equation}
With this function, the entropy production
(\ref{entropyproductionGNEKIN}) has a very simple form:
\begin{equation}\label{EntropProdtheta}
\frac{\D S}{\D t}=V\left.\frac{\D \theta(\lambda)}{\D
\lambda}\right|_{\lambda=1}
\end{equation}

The auxiliary function $\theta(\lambda) $ allows the following
interpretation. Let us introduce the deformed stoichiometric
mechanism with the stoichiometric vectors,
$$\alpha_{\rho}(\lambda)=\lambda
\alpha_{\rho}+(1-\lambda)\beta_{\rho}\, , \;
\beta_{\rho}(\lambda)=\lambda
\beta_{\rho}+(1-\lambda)\alpha_{\rho}$$, which is the initial
mechanism when $\lambda=1$, the inverted mechanism with interchange
of $\alpha$ and $\beta$ when $\lambda=0$, the trivial mechanism (the
left and right hand sides of the stoichiometric equations coincide)
when $\lambda=1/2$.

For the deformed mechanism, let us take the same kinetic factors and
calculate the Boltzmann factors with $\alpha_{\rho}(\lambda)$:
$$r_{\rho}(\lambda)=\varphi_{\rho}\exp(\alpha_{\rho}(\lambda), \check{\mu}
) $$ In this notation, the auxiliary function $\theta(\lambda)$ is a
sum  of reaction rates for the deformed reaction~mechanism:
$$\theta(\lambda)=\sum_{\rho}r_{\rho}(\lambda) $$

In particular, $\theta(1)=\sum_{\rho}r_{\rho}$, this is just the sum
of reaction rates.

Function $\theta(\lambda)$ is convex. Indeed
$$\frac{\D^2 \theta(\lambda)}{\D
\lambda^2}=\sum_{\rho}\varphi_{\rho}(\gamma_{\rho},\check{\mu})^2
\exp[(\check{\mu},(\lambda
\alpha_{\rho}+(1-\lambda)\beta_{\rho}))]\geq 0$$

This convexity gives the following {\em necessary and sufficient
condition for positivity of entropy~production}:
 $$\frac{\D S}{\D t}> 0 \mbox{ if and only if }\theta(\lambda)
 < \theta(1) \mbox{ for some } \lambda < 1 $$

In several next subsections we study various important particular
sufficient conditions for positivity of entropy production.

\subsubsection{6.2.2. Detailed Balance}

The most celebrated condition which gives the positivity of entropy
production is the principle of detailed balance. Boltzmann used this
principle to prove his famous $H$-theorem \cite{Boltzmann}.

Let us join elementary reactions in pairs:
\begin{equation}\label{stoichiomerEqPairs}
\sum_i\alpha_{\rho i}A_i \rightleftharpoons \sum_i \beta_{\rho i}
A_i
\end{equation}
After this joining, the total amount of stoichiometric equations
decreases. If there is no reverse reaction then we can add it
formally, with zero kinetic factor. We will distinguish the reaction
rates and kinetic factors for direct and inverse reactions by the
upper plus or minus:
$$r_{\rho}^+=\varphi_{\rho}^+\exp(\alpha_{\rho},\check{\mu})\, , \;
r_{\rho}^-=\varphi_{\rho}^-\exp(\beta_{\rho},\check{\mu})\, , \;
r_{\rho}=r_{\rho}^+-r_{\rho}^-$$

\begin{equation}\label{generalKINURdetbal}
\frac{\D N}{\D t}=V\sum_{\rho}\gamma_{\rho} r_{\rho}
\end{equation}

In this notation, the principle of detailed balance is very simple:
The thermodynamic equilibrium in the direction $\gamma_{\rho}$,
given by the standard condition $(\gamma_{\rho},\check{\mu})=0$, is
equilibrium for the corresponding pair of mutually reverse reactions
from (\ref{stoichiomerEqPairs}). For kinetic factors this transforms
into the simple and beautiful~condition:
$$\varphi_{\rho}^+\exp(\alpha_{\rho},\check{\mu})=\varphi_{\rho}^-\exp(\beta_{\rho},\check{\mu}) \Leftrightarrow (\gamma_{\rho},\check{\mu})=0$$
therefore
\begin{equation}\label{detailed balance}
\varphi_{\rho}^+=\varphi_{\rho}^-
\end{equation}

For the systems with detailed balance we can take
$\varphi_{\rho}=\varphi_{\rho}^+=\varphi_{\rho}^-$ and write for the
reaction rate:
$$r_{\rho}=\varphi_{\rho}
(\exp(\alpha_{\rho},\check{\mu})-\exp(\beta_{\rho},\check{\mu}))$$
M. Feinberg called this kinetic law the ``Marselin-De Donder''
kinetics \cite{Feinberg1972_a}. This representation of the reaction
rates gives for the auxiliary function $\theta(\lambda)$:
\begin{equation}
\theta(\lambda)=\sum_{\rho}\varphi_{\rho}(\exp[(\check{\mu},(\lambda
\alpha_{\rho}+(1-\lambda)\beta_{\rho}))] +\exp[(\check{\mu},(\lambda
\beta_{\rho}+(1-\lambda)\alpha_{\rho}))])
\end{equation}
Each term in this sum is symmetric with respect to change $\lambda
\mapsto (1-\lambda)$. Therefore, $\theta(1)=\theta(0)$ and, because
of convexity of $\theta(\lambda)$, $\theta'(1)\geq 0$. This means
positivity of entropy production.

The principle of detailed balance is a sufficient but not a
necessary condition of the positivity of entropy production.
This was clearly explained, for example, by L. Onsager
\cite{Onsager1931a,Onsager1931b}. Interrelations between
positivity of entropy production, Onsager reciprocal relations
and detailed balance were analyzed in detail by N.G. van Kampen
\cite{VKampen1973}.

\subsubsection{6.2.3. Complex Balance}

The principle of detailed balance gives us $\theta(1)=\theta(0)$ and
this equality holds for each pair of mutually reverse reactions.

Let us start now from the equality $\theta(1)=\theta(0)$. We return
to the initial stoichiometric equations
(\ref{stoichiometricequation}) without joining the direct and
reverse reactions. The equality reads
\begin{equation}\label{0=1}
\sum_{\rho}\varphi_{\rho}\exp(\check{\mu},\alpha_{\rho})=
\sum_{\rho}\varphi_{\rho}\exp(\check{\mu},\beta_{\rho})
\end{equation}
Exponential functions $\exp(\check{\mu},y)$ form linearly
independent family in the space of functions of $\check{\mu}$ for
any finite set of pairwise different vectors $y$. Therefore, the
following approach is natural: Let us equalize in (\ref{0=1}) the
terms with the same Boltzmann-type factor $\exp(\check{\mu},y)$.
Here we have to return to the complex-based representation of
reactions (see Section~\ref{CoplexStoi}).

Let us consider the family of vectors
$\{\alpha_{\rho},\beta_{\rho}\}$ ($\rho=1, \ldots ,m$). Usually,
some of these vectors coincide. Assume that there are $q$ different
vectors among them. Let $y_1, \ldots, y_q$ be these vectors. For
each $j=1, \ldots, q$ we take
$$R_j^+=\{\rho\, | \,\alpha_{\rho}=y_j\}\, , \; R_j^-=\{\rho\, |
\,\beta_{\rho}=y_j\}$$

We can rewrite the equality (\ref{0=1}) in the form
\begin{equation}\label{0=1incomplex}
\sum_{j=1}^q \exp(\check{\mu},y_j)\left[\sum_{\rho\in
R_j^+}\varphi_{\rho}- \sum_{\rho\in R_j^-}\varphi_{\rho}\right]=0
\end{equation}
The Boltzmann factors $\exp(\check{\mu},y_j)$ form the linearly
independent set. Therefore the natural way to meet these condition
is: For any $j=1, \ldots, q$
\begin{equation}\label{complexbalanceGENKIN}
\sum_{\rho\in R_j^+}\varphi_{\rho}- \sum_{\rho\in
R_j^-}\varphi_{\rho}=0
\end{equation}
This is the general  {\em complex balance condition}. This condition
is sufficient for entropy growth, because it provides the equality
$\theta(1)=\theta(0)$.

If we assume that $\varphi_{\rho}$ are constants or, for chemical
kinetics, depend only on temperature, then the conditions
(\ref{complexbalanceGENKIN}) give the general solution to equation
(\ref{0=1incomplex}).

The complex balance condition is more general than the detailed
balance. Indeed, this is obvious: For the master equation
(\ref{MasterEq}) the complex balance condition is trivially valid
for all admissible constants. The first order kinetics always
satisfies the complex balance conditions. On the contrary, the class
of the master equations with detailed balance is rather special. The
dimension of the class of all master equations has dimension $n^2-n$
(constants for all transitions $A_i\to A_j$ are independent). For
the time-reversible Markov chains (the master equations with
detailed balance) there is only $n(n+1)/2-1$ independent constants:
$n-1$ for equilibrium state and $n(n-1)/2$ for transitions $A_i\to
A_j$ ($i>j$), because for reverse transitions the constant can be
calculated through the detailed balance.

In general, for nonlinear reaction systems, the complex balance
condition is not necessary for entropy growth. In the next section
we will give a more general condition and demonstrate that there are
systems that violate the complex balance condition but satisfy this
more general inequality.

\subsubsection{6.2.4. $G$-Inequality}

Gorban \cite{G1} proposed the following inequality for analysis of
accordance between thermodynamics and kinetics:
$\theta(1)\geq\theta(0)$. This means that for any values of
$\check{\mu}$
\begin{equation}\label{0<1}
\sum_{\rho}\varphi_{\rho}\exp(\check{\mu},\alpha_{\rho})\geq
\sum_{\rho}\varphi_{\rho}\exp(\check{\mu},\beta_{\rho})
\end{equation}
In the form of sum over complexes (similarly to
(\ref{0=1incomplex})) it has the form
\begin{equation}\label{0<1incomplex}
\sum_{j=1}^q \exp(\check{\mu},y_j)\left[\sum_{\rho\in
R_j^+}\varphi_{\rho}- \sum_{\rho\in R_j^-}\varphi_{\rho}\right]\geq
0
\end{equation}
Let us call these inequalities, (\ref{0<1}), (\ref{0<1incomplex}),
the $G$-inequalities.

Here, two remarks are needed. First, functions
$\exp(\check{\mu},y_j)$ are linearly independent but this does not
allow us to transform inequalities (\ref{0<1incomplex}) similarly to
(\ref{complexbalanceGENKIN}) even for constant kinetic factors:
Inequality between linear combinations of independent functions may
exist and the ``simplified system''
$$\sum_{\rho\in R_j^+}\varphi_{\rho}-
\sum_{\rho\in R_j^-}\varphi_{\rho}\geq 0 \mbox{ for all } j $$ is
not equivalent to the $G$-inequality.

Second, this simplified  inequality is equivalent to the complex
balance  condition (with equality instead of $\geq$). Indeed, for
any  $\rho=1, \ldots , m$ there exist exactly one $j_1$ and one
$j_2\neq j_1$ with properties: $\rho \in R_{j_1}^+$, $\rho \in
R_{j_2}^-$. Therefore, for any reaction mechanism with reaction
rates (\ref{GeneralReactionRate}) the identity holds:
$$\sum_{\rho}\left[\sum_{\rho\in R_j^+}\varphi_{\rho}-
\sum_{\rho\in R_j^-}\varphi_{\rho}\right] = 0  $$ If all terms in
this sum are non-negative then all of them are zeros.

Nevertheless, if at least one of the vectors $y_j$ is a convex
combination of others, $$\sum_{k, \, k\neq j}\lambda_k y_k=y_j
\mbox{ for some } \lambda_k \geq 0, \, \sum_{k, \, k\neq j}
\lambda_k=1$$
 then the $G$ inequality has more solutions than the condition
 of complex balance. Let us take a very simple example with two
 components, $A_1$ and $A_2$, three reactions and three
 complexes: $$2A_1 \rightleftharpoons A_1+A_2, \, 2A_2
 \rightleftharpoons A_1+A_2, \, 2A_1\rightleftharpoons 2A_2$$
 $$y_1=(2,0), \, y_2=(0,2), \, y_3=(1,1)\, ,$$
  $$R_1^+=\{1,3\}, \, R_2^+=\{2,-3\}, \,   R_3^+=\{-1,-2\} $$
  $$R_1^-=\{-1,-3\}, \,   R_2^-=\{-2,3\}, \,  R_3^-=\{1,2\} $$
The complex balance condition for this system is
\begin{equation}\label{ComplexBalanceSimpleEx}
\begin{split}
&(\varphi_1-\varphi_{-1})+(\varphi_3-\varphi_{-3})=0 \, \\
&(\varphi_2-\varphi_{-2})-(\varphi_3-\varphi_{-3})=0
\end{split}
\end{equation}
The $G$-inequality for this system is
\begin{equation}\label{G-ineqExample}
\begin{split}
&(\varphi_1+\varphi_3-\varphi_{-1}-\varphi_{-3})a^2+(\varphi_2+\varphi_{-3}-
\varphi_{-2}-\varphi_3)b^2 \\
&+(\varphi_{-1}+\varphi_{-2}-\varphi_1-\varphi_2)ab\geq 0 \mbox{ for
all } a,b>0
\end{split}
\end{equation}
(here, $a,b$ stand for $\exp(\check{\mu}_1), \,
\exp(\check{\mu}_2)$). Let us use for the coefficients at $a^2$ and
$b^2$ notations $\psi_a$ and $\psi_b$. Coefficient at $ab$ in
(\ref{G-ineqExample}) is $-(\psi_a+\psi_b)$, linear combinations
$\psi_a=\varphi_1+\varphi_3-\varphi_{-1}-\varphi_{-3}$ and
$\psi_b=\varphi_2+\varphi_{-3}- \varphi_{-2}-\varphi_3$ are linearly
independent functions of variables $\varphi_i$ ($i=\pm 1, \pm 2, \pm
3$) and we get the following task: To find all pairs of numbers
$(\psi_a,\psi_b) \in \mathbb{R}^2$ which satisfy the inequality
$$\psi_a a^2+ \psi_b b^2\geq (\psi_a+\psi_b) ab\; \mbox{ for all
} a,b>0 $$ Asymptotics $a\to 0$ and $b\to 0$ give $\psi_a,\psi_b\geq
0$.

Let us use homogeneity of functions in (\ref{G-ineqExample}),
exclude one normalization factor from $a,b$ and one factor from
$\psi_a,\psi_b$ and reduce the number of variables: $b=1-a$,
$\psi_a=1-\psi_b$: We have to find all such $\psi_b\in[0,1]$ that
for all $a\in ]0,1[$
$$a^2(1-\psi_b)+(1-a)^2\psi_b-a(1-a)\geq 0$$
The minimizer of this quadratic function of $a$ is
$a_{\min}=\frac{1}{4}+\frac{1}{2}\psi_b$, $a_{\min}\in ]0,1[$ for
all $\psi_b\in[0,1]$. The minimal value is
$-2(\frac{1}{2}\psi_b-\frac{1}{4})^2$. It is nonnegative if and only
if $\psi_b=\frac{1}{2}$. When we return to the non-normalized
variables $\psi_a,\psi_b$ then we get the general solution of the
$G$-inequality for this example: $\psi_a=\psi_b\geq0$. For the
kinetic factors this means:
\begin{equation}\label{GIneqSimple}
\begin{split}
(\varphi_1-\varphi_{-1})+2(\varphi_3-\varphi_{-3})- (\varphi_2-
\varphi_{-2})=0\, \\
(\varphi_1-\varphi_{-1})+(\varphi_3-\varphi_{-3})\geq0\, \\
(\varphi_2- \varphi_{-2}) - (\varphi_{3}-\varphi_{-3})\geq0
\end{split}
\end{equation}
These conditions are wider (weaker) than the complex balance
conditions for this example (\ref{ComplexBalanceSimpleEx}).

In the Stueckelberg language \cite{Stueckelberg1952}, the
microscopic reasons for the $G$-inequality instead of the complex
balance (\ref{kinetic multiplier complex balance}) can be explained
as follows: Some channels of the scattering are unknown (hidden),
hence, instead of unitarity of $S$-matrix (conservation of the
microscopic probability) we have an inequality (the microscopic
probability does not increase).

We can use other values of $\lambda_0\in[0,1[$ in inequality
$\theta(1)\geq \theta(\lambda_0)$ and produce constructive
sufficient conditions of accordance between thermodynamics and
kinetics. For example, condition $\theta(1)\geq \theta(1/2)$ is
weaker than $\theta(1)\geq \theta(0)$ because of convexity
$\theta(\lambda)$.

One can ask a reasonable question: Why we do not use directly
positivity of entropy production ($\theta'(1)\geq 0$) instead of
this variety of sufficient conditions. Of course, this is possible,
but inequalities like $\theta(1)\geq \theta(0)$ or equations like
$\theta(1)= \theta(0)$ include linear combinations of exponents of
linear functions and often can be transformed in algebraic equations
or inequalities like in the example above. Inequality
$\theta'(1)\geq 0$ includes transcendent functions like $f\exp f$
(where $f$ is a linear function) which makes its study more
difficult.

\section{Linear Deformation of Entropy \label{Sec:EntropyDeform}}

\subsection{If Kinetics Does not Respect Thermodynamics then
Deformation of Entropy May Help}

Kinetic equations in the general form (\ref{generalKINUR}) are very
general, indeed. They can be used for the approximation of any
continuous time dynamical system on compact $U$ \cite{Ocherki}. In
previous sections we demonstrated how to construct the system in the
form (\ref{generalKINUR}) with positivity of the entropy production
when the entropy function is given.

Let us consider a reverse problem.  Assume that a system in the form
(\ref{generalKINUR}) is given but the entropy production is not
always positive. How to find a new entropy function for this system
to guarantee the positivity of entropy production?

Existence of such an entropy is very useful for analysis of
stability of the system. For example, let us take an arbitrary Mass
Action Law system (\ref{MALreaction rate}). This is a rather general
system with the polynomial right hand side. Its stability or
instability is not obvious a priori. It is necessary to check
whether bifurcations of steady states, oscillations and other
interesting effects of dynamics are possible for this system.

With the positivity of entropy productions these questions are much
simpler (for application of thermodynamic potentials to stability
analysis see, for example,
\cite{Yablonskii1991,Ocherki,VolpertKhudyaev1985,HangosAtAl1999}).
If $\D S /\D t \geq 0$ and it is zero only in steady states, then
any motion in compact $U$ converges to a steady state and all the
non-wandering points are steady states. (A non-wandering point is
such a point $x \in U$ that for any $T>0$ and $\varepsilon > 0$
there exists such a motion $c(t)\in U$ that (i)
$\|c(0)-x\|<\varepsilon$ and (ii) $\|c(T')-x\|<\varepsilon$ for some
$T'>T$: A motion returns in an arbitrarily small vicinity of $x$
after an arbitrarily long time.) Moreover, the global maximizer of
$S$ in $U$ is an asymptotically stable steady state (at least,
locally). It is a globally asymptotically stable point if there is
no other steady state in $U$.

For the global analysis of an arbitrary system of differential
equations, it is desirable either to construct a general Lyapunov
function or to prove that it does not exist. For the Lyapunov
functions of the general form this task may be quite difficult.
Therefore, various finite-dimensional spaces of trial functions are
often in use. For example, quadratic polynomials of several
variables provide a very popular class of trial Lyapunov function.

In this section, we discuss the $n$-parametric families of Lyapunov
functions which are produced by the addition of linear function to
the entropy:
\begin{equation}\label{EntropyLinearDeformation}
S(N)\mapsto S_{\Delta \check{\mu}}(N)=S(N)-\sum_i \Delta
\check{\mu}_i N_i
\end{equation}
The change in potentials $\check{\mu}$ is simply the addition of
$\Delta \check{\mu}$: $\check{\mu}_i \mapsto \check{\mu}_i + \Delta
\check{\mu}_i$.

Let us take a general kinetic equation (\ref{generalKINUR}). We are
looking for a transformation that does not change the reaction
rates. The Boltzmann factor $\Omega_{\rho}=\exp(\check{\mu},
\alpha_{\rho})$ transforms due to the change of the entropy:
$\Omega_{\rho} \mapsto \Omega_{\rho} \exp(\Delta \check{\mu},
\alpha_{\rho})$. Therefore, to preserve the reaction rate, the
transformation of the kinetic factors should be $\varphi_{\rho}
\mapsto \varphi_{\rho} \exp(\Delta \check{\mu}, \alpha_{\rho})$ in
order to keep the product $r_{\rho}=\Omega_{\rho} \varphi_{\rho}$
constant.

For the new entropy, $S=S_{\Delta \check{\mu}}$, with the new
potential and kinetic factors, the entropy production is given by
(\ref{entropyproductionGNEKINseformed}):

\begin{equation}\label{entropyproductionGNEKINseformed}
\begin{split}
\frac{\D S}{\D t}=&-\sum_{\rho }(\gamma_{\rho },
\check{\mu})\varphi_{\rho}\exp( \alpha_{\rho }, \check{\mu})\\=
&-\sum_{\rho }(\gamma_{\rho }, \check{\mu}^{\rm old}+\Delta
\check{\mu})\varphi_{\rho}^{\rm old} \exp( \alpha_{\rho },
\check{\mu}^{\rm old})\\ =&\frac{\D S^{\rm old}}{\D t}-
 \sum_{\rho }(\gamma_{\rho }, \Delta \check{\mu})\varphi_{\rho}^{\rm old}
\exp( \alpha_{\rho }, \check{\mu}^{\rm old})\,
\end{split}
\end{equation}
where the superscript ``old'' corresponds to the non-deformed
quantities.

\subsection{Entropy Deformation for Restoration of Detailed Balance}

It may be very useful to find such a vector $\Delta \check{\mu}$
that in new variables $\varphi_{\rho}^+=\varphi_{\rho}^-$. For the
analysis of the detailed balance condition, we group reactions in
pairs of mutually inverse reactions (\ref{stoichiomerEqPairs}). Let
us consider an equation of the general form
(\ref{generalKINURdetbal}) with $r_{\rho}=r_{\rho}^+-r_{\rho}^-$,
$\varphi_{\rho}^{\pm}>0$.

The problem is: To find such a vector $\Delta \check{\mu}$ that
\begin{equation}\label{restoreDetBalan}
\varphi_{\rho}^+\exp{(\Delta \check{\mu}, \alpha_{\rho})}=
\varphi_{\rho}^-\exp{(\Delta \check{\mu}, \beta_{\rho})}
\end{equation}
or, in the equivalent form of the linear equation
\begin{equation}\label{restoreDetBalanLinearEq}
(\Delta \check{\mu}, \gamma_{\rho})= \ln
\left(\frac{\varphi_{\rho}^+}{\varphi_{\rho}^-}\right)
\end{equation}

The necessary and sufficient conditions for the existence of such
$\Delta \check{\mu}$ are known from linear algebra: For every set of
numbers $a_{\rho}$ (${\rho}=1, \ldots ,m$)
\begin{equation}\label{detBalRestCond}
\sum_{\rho} a_{\rho} \gamma_{\rho}=0 \Rightarrow \sum_{\rho}
a_{\rho} \ln
\left(\frac{\varphi_{\rho}^+}{\varphi_{\rho}^-}\right)=0
\end{equation}
To check these conditions, it is sufficient to find a basis of
solutions of the uniform systems of linear~equations
$$\sum_{\rho} a_{\rho} \gamma_{\rho i}=0 \;\;(i=1, \ldots ,
m)$$ (that is, to find a basis of the left kernel of the matrix
$\Gamma$, ${\rm coim} \Gamma$, where $\Gamma = (\gamma_{\rho i})$)
and then check for these basis vectors the condition $\sum_{\rho}
a_{\rho} \ln
\left(\frac{\varphi_{\rho}^+}{\varphi_{\rho}^-}\right)=0$ to prove
or disprove that the vector with coordinates
$\ln\left(\frac{\varphi_{\rho}^+}{\varphi_{\rho}^-}\right)$ belongs
to the image of $\Gamma$, ${\rm im} \Gamma$.

For some of the reaction mechanisms it is possible to restore the
detailed balance condition for the general kinetic equation
unconditionally. For these reactions, for any set of positive
kinetic factors, there exists such a vector $\Delta \check{\mu}$
that the detailed balance condition (\ref{restoreDetBalanLinearEq})
is valid for the deformed entropy. According to
(\ref{detBalRestCond}) this means that there is no nonzero solution
$a_{\rho}$ for  the equation $\sum_{\rho} a_{\rho} \gamma_{\rho}=0$.
In other words, vectors $\gamma_{\rho}$ are independent.

\subsection{Entropy Deformation for Restoration of Complex Balance  \label{Sec:ComplexBalanceDeformEnt}}

The complex balance conditions (\ref{complexbalanceGENKIN}) are, in
general, weaker than the detailed balance but they are still
sufficient for the entropy growth.

Let us consider an equation of the general form
(\ref{generalKINURdetbal}). We need to find such a vector $\Delta
\check{\mu}$ that in new variables with the new entropy and kinetic
factors the complex balance conditions \linebreak $\sum_{\rho\in
R_j^+}\varphi^{\rm new}_{\rho}- \sum_{\rho\in R_j^-}\varphi^{\rm
new}_{\rho}=0$ hold.

For our purpose, it is convenient to return to the presentation of
reactions as transitions between complexes. The complexes,
$\Theta_1, \ldots, \Theta_q$ are the linear combinations,
$\Theta_j=(y_j,A)$.

Each elementary reaction (\ref{stoichiometricequation}) with the
reaction number $\rho$ may be represented in the form $\Theta_j \to
\Theta_l$, where $\Theta_{j}=\sum y_j A_j$, $\rho\in  R_j^+$
($\alpha_{\rho}=y_j$)  and $\rho\in  R_j^-$ ($\beta_{\rho}=y_l$).
For this reaction, let us use the notation
$\varphi_{\rho}=\varphi_{lj}$. We used this notation in the analysis
of kinetics of compounds (Section~\ref{Sec:QEelimComp}). The complex
balance conditions are
\begin{equation}\label{complexbalanceComplex}
\sum_{j,\,j\neq l}(\varphi_{lj}-\varphi_{jl})=0
\end{equation}
To obtain these conditions after the entropy deformation, we have to
find such $\Delta \check{\mu}$ that
\begin{equation}\label{complexbalanceComplexDeform}
\sum_{j,\,j\neq l}(\varphi_{lj}\exp{(\Delta \check{\mu},y_j)
}-\varphi_{jl}\exp{(\Delta \check{\mu},y_l) })=0
\end{equation}
This is exactly the equation for equilibrium of a Markov chain with
transition coefficients $\varphi_{lj}$. Vector $(\Delta
\check{\mu},y_j)$ should be an equilibrium state for this chain
(without normalization to the unit sum of~coordinates).

For this finite Markov chain a graph representation is useful:
Vertices are complexes and oriented edges are reactions. To provide
the existence of a positive equilibrium we assume {\em weak
reversibility} of the chain: If there exists an oriented path from
$\Theta_j$ to $\Theta_l$ then there exists an oriented path from
$\Theta_l$ to $\Theta_j$.

Let us demonstrate how to transform this problem of entropy
deformation into a linear algebra problem. First of all, let us find
any positive equilibrium of the chain, $\varsigma^*_j>0$:
\begin{equation}\label{complexChainEquilibrium}
\sum_{j,\,j\neq
l}(\varphi_{lj}\varsigma^*_j-\varphi_{jl}\varsigma^*_l)=0
\end{equation}
This is a system of linear equations. If we have already an
arbitrary equilibrium of the chain then other equilibria allow a
very simple description. We already found this description for
kinetics of compounds~(\ref{compoundEq})

Let us consider the master equation for the Markov chain with
coefficients $\varphi_{lj}$ and apply the formalism from Appendix 2:
\begin{equation}\label{MarkovComplexChainMaster}
\frac{\D \varsigma}{\D t}=\sum_{j,\,j\neq
l}(\varphi_{lj}\varsigma_j-\varphi_{jl}\varsigma_l)=0
\end{equation}

Let the graph of complex transformations $\Theta_j \to \Theta_l$
have $d$ connected components $C_s$ and let $V_s$ be the set of
indexes of those $\Theta_j$ which belong to $C_s$: $\Theta_j \in
C_s$ if and only if $j \in V_s$. For each $C_s$ there exists a
conservation law $\beta_s(\varsigma)$ for the master equation
(\ref{LinearConservation}), $\beta_s(\varsigma)=\sum_{j\in V_s}
\varsigma_j$.

For any set of positive values of $\beta_s$ ($s=1, \ldots , q$)
there exists a unique equilibrium vector $\varsigma^{\rm eq}$
\linebreak for (\ref{MarkovComplexChainMaster}) with this values
$\beta_s$ (\ref{compoundEq}),  (\ref{compoundEqA}). The set of
equilibria is a linear space with the natural coordinates $\beta_s$
($s=1, \ldots ,d$). We are interested in the positive orthant of
this space, $\beta_s>0$. For positive $\beta_s$, logarithms of
$\varsigma^{\rm eq}$ form a $d$-dimensional linear manifold in $R^q$
(\ref{complexMarkovEq1A}). The natural coordinates on this manifold
are $\ln \beta_s$.

Let us notice that the vector $\varsigma^{\circ}$ with coordinates
$$\varsigma^{\circ}_j=\left(\frac{\varsigma_j^*(c,T)}{\sum_{l\in V_s
}\varsigma_l^*(c,T)} \right)\mbox{ for } j\in V_s$$ is also an
equilibrium for (\ref{MarkovComplexChainMaster}). This equilibrium
is normalized to unit values of all $\beta_s(\varsigma^{\circ})$. In
the coordinates $\ln \beta_s$ this is the origin.  The equations for
$\Delta \check{\mu}$ are
\begin{equation}\label{RestoreComplBalanceEq}
(\Delta \check{\mu},y_j)-\ln \beta_s = \ln \varsigma^{\circ}_j
\mbox{ for } j\in V_s
\end{equation}
This is a system of linear equations with respect to $n+d$ variables
$\Delta \check{\mu}_i$ ($i=1,\ldots , n$) and \linebreak $\ln
\beta_s$ ($s=1, \ldots , d$). Let the coefficient matrix of this
system be denoted by $\mathbf{M}$.

Analysis of solutions and solvability of such equations is one of
the standard linear algebra tasks. If this system has a solution
then the complex balance in the original system can be restored by
the linear deformation of the entropy. If this system is solvable
for any right hand side, then for this reaction mechanism we always
can find the entropy, which provides the complex balance condition.

Unconditional solvability of (\ref{RestoreComplBalanceEq}) means
that the left hand side matrix of this system has rank $q$. Let us
express this rank through two important characteristics: It is ${\rm
rank}\{\gamma_1, \ldots , \gamma_m\}+d$, where $d$ is the number of
connected components in the graph of transformation of complexes.

To prove this formula, let us write down the matrix $\mathbf{M}$ of
the system (\ref{RestoreComplBalanceEq}). First, we change the
enumeration of complexes. We group the complexes from the same
connected component together and arrange these groups in the order
of the connected component number. After this change of enumeration,
$\{1, \ldots , |V_1|\} = V_1$, $\{|V_1|+1, \ldots , |V_1|+|V_2| \}=
V_2$, ..., $\{|V_1|+|V_2|+\ldots + |V_{d-1}|+1, \ldots ,
|V_1|+|V_2|+\ldots + |V_{d}|\}= V_d$.

Let $y_j$ be here the row vector.  The matrix is
\begin{equation}\label{matrixCOmplexBalance}
\mathbf{M}=\left[\begin{array}{lcccc}
y_1&1&0&\ldots&0\\
\vdots&\vdots&\vdots&\vdots&\vdots\\
y_{|V_1|}&1&0&\ldots&0\\
y_{|V_1|+1}&0&1&\ldots&0\\
\vdots&\vdots&\vdots&\vdots&\vdots\\
y_{|V_1|+\ldots+|V_{d}|}&0&0&\ldots&1
\end{array}\right]
\end{equation}
$\mathbf{M}$ consists of $d$ blocks $\mathbf{M}_s$, which correspond
to connected components $C_s$ of the graph of transformation of
complexes:
\begin{equation}\label{matrixBlockCOmplexBalance}
\mathbf{M}_s\left[\begin{array}{lcccc}
y_{|V_1|+\ldots+|V_{s-1}|+1}&0&\ldots&1&\ldots\\
\vdots&\vdots&\vdots&\vdots&\vdots\\
y_{|V_1|+\ldots+|V_{s}|}&0&\ldots&1&\ldots
\end{array}\right]
\end{equation}
The first $n$ columns in this matrix are filled by the vectors $y_j$
of complexes, which belong to the component $C_s$,  then follow
$s-1$ columns of zeros, after that, there is one column of units,
and then again zeros. Here, in (\ref{matrixCOmplexBalance}),
(\ref{matrixBlockCOmplexBalance}) we multiplied the last $d$
columns by $-1$. This operation does not change the rank of the
matrix.

Other elementary operations that do not change the rank are: We can
add to any row (column) a linear combination of other rows
(columns).

We will use these operations to simplify blocks
(\ref{matrixBlockCOmplexBalance}) but first we have to recall
several properties of spanning trees \cite{WuChao2004}. Let us
consider a connected, undirected graph $G$ with the set of vertices
$\mathcal{V}$ and the set of edges $\mathcal{E}\subset
\mathcal{V}\times \mathcal{V}$. A spanning tree of $G$ is a
selection of edges of $G$ that form a tree spanning every vertex.
For a connected graph with $V$ vertices, any spanning tree has $V-1$
edges. Let for each vertex $\Theta_j$ of $G$ a $n$-dimensional
vector $y_i$ is given. Then for every edge $(\Theta_j,\Theta_l)\in
\mathcal{E}$ a vector $\gamma_{jl}=y_{j}-y_l$ is defined. We
identify vectors $\gamma$ and $-\gamma$ and the order of $j,l$ is
not important. Let us use $\Gamma_{G}$ for this \linebreak set of
$\gamma_{jl}$: $$\Gamma_{G}=\{y_{j}-y_l\, | \,(\Theta_j,\Theta_l)\in
\mathcal{E}\}$$ For any spanning tree $T$ of graph $G$ we have the
following property:
\begin{equation}\label{spanTreeRank}
{\rm span} \Gamma_{G}= {\rm span} \Gamma_{T}
\end{equation}
in particular, ${\rm rank} \Gamma_{G} = {\rm rank} \Gamma_T$.

For the digraphs of reactions between complexes, we create
undirected graphs just by neglecting the directions of edges. We
keep for them the same notations as for original digraphs. Let us
select any spanning tree $T_s$ for the connected component $C_s$ in
the graph of transformation of complexes. In $T_s$ we select
arbitrarily a root complex. After that, any other complex $\Theta_j$
in $C_s$ has a unique parent. This is the vertex connected to it on
the path to the root. For the root complex of $C_s$ we use special
notation $\Theta_s^{\circ}$.

Now, we transform the block (\ref{matrixBlockCOmplexBalance})
without change of rank: For each non-root complex we subtract
from the corresponding row the row which correspond to its
unique parent. After these transformations (and, may be, some
permutations of rows), the block $\mathbf{M}_s$ get the
following form:
\begin{equation}
\left[\begin{array}{lcccc}
\gamma^s_1&0&\ldots&0&\ldots\\
\gamma^s_2&0&\ldots&0&\ldots\\
\vdots&\vdots&\vdots&\vdots&\vdots\\
\gamma^s_{|V_1|-1}&0&\ldots&0&\ldots\\
y_s^{\circ}&0&\ldots&1&\ldots
\end{array}\right]
\end{equation}
Here, $\{\gamma^s_1,\gamma^s_2, \ldots , \gamma^s_{|V_1|-1}\}$ is
$\Gamma_{T_s}$ for the spanning tree $T_s$ and $y_s^{\circ}$ is the
coefficient vector for the root complex $\Theta_s^{\circ}$.

From the obtained structure of blocks we immediately find that the
rank of the rows with $\gamma$ is ${\rm rank}\{\gamma_1, \ldots ,
\gamma_m\}+d$ due to (\ref{spanTreeRank}). Additional $d$ rows with
$y_s^{\circ}$ are independent due to their last coordinates and  add
$d$ to rank. Finally
\begin{equation}\label{rankM}
{\rm rank} \mathbf{M}={\rm rank}\{\gamma_1, \ldots , \gamma_m\}+d
\end{equation}
Obviously, ${\rm rank} \mathbf{M}\leq q$.

In particular, from the formula (\ref{rankM}) immediately follows
the description of the reaction mechanisms, for which it is always
possible to restore the thermodynamic properties by the linear
deformation of the entropy.

\vspace{2mm}\noindent{\bf The deficiency zero theorem.} {\it If
${\rm rank} \mathbf{M}= q$ then it is always possible to restore the
positivity of the entropy production by the linear deformation of
the entropy.} \vspace{2mm}

Feinberg \cite{Feinberg1972} called the difference $q-{\rm rank}
\mathbf{M}$ the {\em deficiency} of the reaction network. For
example, for the ``Michaelis-Menten'' reaction mechanism
$E+S\rightleftharpoons ES \rightleftharpoons P+S$ ${\rm
rank}\{\gamma_1,  \gamma_2\}=2$, $d=1$, $q=3$, ${\rm rank}
\mathbf{M}=3$ and deficiency is 0.

For the adsorption (the Langmuir-Hinshelwood) mechanism of CO
oxidation (\ref{Langmuir-Hinshelwood}) ${{\rm rank}\{\gamma_1,
\gamma_2, \gamma_3 \}~=~3}$, $d=3$, $q=6$, ${\rm rank} \mathbf{M}=6$
and deficiency is 0. To apply the results about the entropy
deformation to this reaction mechanism, it is necessary to introduce
an inverse reaction to the third elementary reaction in
(\ref{Langmuir-Hinshelwood}), PtO+PtCO$\to$CO$_2$+2Pt with an
arbitrarily small but positive constant in order to make the
mechanism weakly reversible.

Let us consider the Langmuir-Hinshelwood mechanism for reduced list
of components. Let us assume that the gas concentrations are
constant because of control or time separation or just as a model
``fast'' system and just include them in the reaction rate constants
for intermediates. Then the mechanism is
2Pt$\rightleftharpoons$2PtO, Pt$\rightleftharpoons$PtCO,
PtO+PtCO$\to$2Pt. For this system, ${\rm rank}\{\gamma_1, \gamma_2,
\gamma_3 \}=2$, $d=2$, $q=5$, ${\rm rank} \mathbf{M}=4$ and
deficiency is 1. Bifurcations in this system are known
\cite{Yablonskii1991}.

For the fragment of the reaction mechanism of the hydrogen
combustion (\ref{hyroexample}), ${\rm rank}\{\gamma_1, \ldots ,
\gamma_m\}=6$, $d=7$, $q=16$, ${\rm rank} \mathbf{M}=6+7=13$
and deficiency is 3.

\subsection{Existence of Points of Detailed and Complex Balance}

Our formulation of the conditions of detailed and complex balance is
not standard: We formulate them as the identities (\ref{detailed
balance}) and (\ref{0=1}). These identities have a global nature and
describe the properties of reaction rates for all states.

The usual approach to the principle of detailed balance is based on
equilibria. The standard formulation is: {\em In all equilibria
every process is balanced with its reverse process}. Without special
forms of kinetic law this principle cannot have any consequences for
global dynamics. This is a trivial but not widely known fact.
Indeed, let a system $\dot{c}=F(c)$ be given in a domain $U \subset
\mathbb{R}^n$ and $\gamma_1, \ldots , \gamma_n$ is an arbitrary
basis in $\mathbb{R}^n$. In this basis, we can always write:
$F(c)=\sum_{\rho=1}^n r_{\rho}(c) \gamma_{\rho}$. For any
equilibrium $c^*$, $r_{\rho}(c^*)=0$. All the ``reaction rates''
$r_{\rho}(c)$ vanish simultaneously. This ``detailed balance'' means
nothing for dynamics because $F$ is an arbitrary vector field. Of
course, if the system of vectors $\{\gamma_{\rho}\}$ is not a basis
but any complete system of vectors then such ``detailed balance''
conditions, $r_{\rho}(c^*)=0$, also do not imply any specific
features of dynamics without special hypotheses about functions
$r_{\rho}(c)$.

Nevertheless, if we fix the kinetic law then the consequences may be
very important. For example, if kinetics of elementary reactions
follow the Mass Action law then the existence of a positive
equilibrium with detailed balance implies existence of the Lyapunov
function in the form of the perfect free entropy:
$$Y=-\sum_i c_i \left(\ln \left(\frac{c_i }{c^*_i}\right)-1\right) $$
where $c_i^*$ is that positive equilibrium with detailed balance
(see, for example, \cite{Yablonskii1991}).

In this section we demonstrate that for the general kinetic law
(\ref{GeneralReactionRate}), which gives the expression of reaction
rates through the entropy gradient, if the kinetic factors are
constant (or a function of temperature) then the existence of the
points of detailed (or complex) balance means that the linear
deformation of the entropy exists which restores the global detailed
(or complex) balance conditions~(\ref{detailed balance}) (or
(\ref{0=1})).

The condition that the kinetic factors are constant means that for a
given set of values $\{\varphi_{\rho}\}$ a state with any admissible
values of $\check{\mu}$ is physically possible (admissible). This
condition allows us to vary the potentials $\check{\mu}$
independently of $\{\varphi_{\rho}\}$.

Let us assume that for the general kinetic system with the
elementary reaction rates given by (\ref{GeneralReactionRate}) a
point of detailed balance exists. This means that for some value of
$\check{\mu}=\check{\mu}^*$ (the detailed balance point in the
Legendre transform) and for all ${\rho}$ $r_{\rho}^+=r_{\rho}^-$:
$$\varphi_{\rho}^+\exp(\alpha_{\rho}, \check{\mu}^* )=\varphi_{\rho}^-\exp(\alpha_{\rho}, \check{\mu}^* )$$
This formula is exactly the condition (\ref{restoreDetBalan}) of
existence of $\Delta \check{\mu}$ which allow us to deform the
entropy for restoring the detailed balance in the global form
(\ref{detailed balance}).

If we assume that the point of complex balance exists then there
exists such a value of $\check{\mu}=\check{\mu}^*$ (a point of
complex balance in the Legendre transform) that
$$\sum_{j,\,j\neq l}(\varphi_{lj}\exp{( \check{\mu}^*,y_j) }-\varphi_{jl}\exp{( \check{\mu}^*,y_l) })=0 $$
This is exactly the deformation condition
(\ref{complexbalanceComplexDeform}) with $\Delta
\check{\mu}=\check{\mu}^*$.

To prove these statements we used an additional condition about
possibility to vary $\check{\mu}$ under \linebreak given
$\{\varphi_{\rho}\}$.

So, we demonstrated that for the general kinetic law
(\ref{GeneralReactionRate}) the existence of a point of detailed
balance is equivalent to the existence of such linear deformation of
the entropy that the global condition (\ref{detailed balance})
holds. Analogously, the existence of a point of complex balance is
equivalent to the global condition of complex balance after some
linear deformation of the entropy.

\subsection{The Detailed Balance is Needed More Often than the Complex Balance\label{DetMicro}}

The complex balance conditions are mathematically nice and more
general than the principle of the detailed balance. They are linked
by Stueckelberg to the Markov models (``$S$-matrix models'') of
microscopic kinetics. Many systems satisfy these conditions (after
linear deformation of the entropy) just because of the algebraic
structure of the reaction mechanism (see
Section~\ref{Sec:ComplexBalanceDeformEnt}). Nevertheless, it is used
much less than the classical detailed balance.

The reason for the rare use of complex balance is simple: It is
less popular because the stronger condition, the principle of
detailed balance, is valid for most of physical and chemical
systems. Onsager revealed the physical reason for detailed
balance \cite{Onsager1931a,Onsager1931b}. This is {\em
microreversibility}: The microscopic laws of motion are
invertible in time: If we observe the microscopic dynamics of
particles in the backward movie then we cannot find the
difference from the real world. This difference occurs in the
macroscopic~world.

In microphysics and the $S$-matrix theory this microreversibility
property has the name ``$T$-invariance''.

Let us demonstrate how $T$-invariance in micro-world implies
detailed balance in macro-world.

Following Gibbs, we accept the ensemble-based point of view on the
macroscopic states: They are probability distributions in the space
of detailed microscopic states.

First of all, we assume that under given values of conservation laws
equilibrium state exists and is~unique.

Second assumption is that the rates of elementary processes are
microscopically observable quantities. This means that somebody (a
``demon''), who observes all the events in the microscopical world
can count the rates of elementary reactions.

Because of $T$-invariance and uniqueness of equilibrium, the
equilibrium is $T$-invariant: If we change all the microscopic time
derivatives (velocities) $v$ to $-v$ then nothing will change.

$T$-transformation changes all reactions to the reverse reactions,
just by reversion of arrows, but the number of the events remains
the same: Any reaction transforms into its reverse reaction but does
not change the reaction rate. This can be formulated also as
follows: $T$-transformation maps all $r_{\rho}^+$ into the
corresponding $r_{\rho}^-$.

Hence, because of the $T$-invariance, the equilibrium rate of each
reaction is equal to the equilibrium rate of the reverse reaction.

The violation of uniqueness of equilibrium for given values of
conservation laws seems improbable. Existence of several equilibria
in thermodynamics is quite unexpected for homogeneous systems but
requires more attention for the systems with phase separation.
Nevertheless, if we assume that a multi-phase system consists of
several homogeneous phases, and each of these phases is in uniform
equilibrium, then we return to the previous assumption (with some
white spots for non-uniform~interfaces).

$T$-invariance may be violated if the microscopic description is not
reversible in time. Magnetic field and the Coriolis force are the
classical examples for violation of the microscopic reversibility.
In a linear approximation near equilibrium the corresponding
modification of the Onsager relations give the Onsager-Casimir
relations \cite{Casimir1945}. There are several attempts for
nonlinear formulation of the Onsager-Casimir relations (see
\cite{Grmela1993}).

The principle of detailed balance seems to be still the best
nonlinear version of the Onsager relations for $T$-invariant
systems, and the conditions of the complex balance seem to give the
proper relations between kinetic coefficients in the absence of the
microscopic reversibility for nonlinear systems. It is important to
mention here that all these relations are used together with the
general kinetic law (\ref{GeneralReactionRate}).

Observability of the rates of elementary reactions deserves a
special study. Two approaches to the reaction rate are possible. If
we accept that the general kinetic law (\ref{GeneralReactionRate})
is valid then we can find the kinetic factors by observation of $\D
c / \D t$ in several points because the Boltzmann factors are
linearly independent. In this sense, they are observable but one can
claim the approximation point of view and state that the general
kinetic law (\ref{GeneralReactionRate}) without additional
conditions on kinetic factors is very general and allows to
approximate any dynamical system. From this point of view, kinetic
coefficients are just some numbers in the approximation algorithm
and are not observable. This means that there is no such a
microscopic thing as the rate of elementary reaction, and the set of
reactions serves just for the approximation of the right hand side
of the kinetic equation. We cannot fully disprove this point of view
but can just say that in some cases the collision-based approach
with physically distinguished elementary reactions is based on the
solid experimental and theoretical background. If the elementary
reactions physically exist then the detailed balance for
$T$-invariant systems is proved.

\section{Conclusions}

We present the general formalism of the Quasiequilibrium
approximation (QE) with the proof of the persistence of entropy
production in the QE approximation (Section~\ref{General}).

We demonstrate how to apply this formalism to chemical kinetics and
give several examples for the Mass Action law kinetic equation. We
discuss the difference between QE and Quasi-Steady-State (QSS)
approximations and analyze the classical Michaelis-Menten and
Briggs-Haldane model reduction approaches
(Section~\ref{MMConfusion}). After that, we use ideas of Michaelis,
Menten and Stueckelberg to create a general approach to kinetics.

Let us summarize the main results of our discussion. First of all,
we believe that this is the finish of the
Michaelis-Menten-Stueckelberg program. The approach to modeling of
the reaction kinetics proposed by Michaelis and Menten in 1913
\cite{MichaelisMenten1913} for enzyme reactions was independently in
1952 applied by Stueckelberg \cite{Stueckelberg1952} to the
Boltzmann equation.

The idea of the complex balance (cyclic balance) relations was
proposed by Boltzmann as an answer to the Lorentz objections against
Boltzmann's proof of the $H$-theorem. Lorentz stated that the
collisions of the polyatomic molecules may have no inverse
collisions. Cercignani and Lampis \cite{CercignaniLamp1981}
demonstrated that the Boltzmann $H$-theorem based on the detailed
balance conditions is valid for the polyatomic molecules under the
microreversibility conditions and this new Boltzmann's idea was not
needed. Nevertheless, this seminal idea was studied further by many
authors \cite{Heitler1944,Coester1951,Watanabe1955} mostly for
linear systems. Stueckelberg \cite{Stueckelberg1952} proved these
conditions for the Boltzmann equation. He used in his proof the
$S$-matrix representation of the micro-kinetics.

Some consequences of the Stueckelberg approach were rediscovered for
the Mass Action law kinetics by Horn and Jackson in 1972
\cite{HornJackson1972} and supplemented by the ``zero deficiency
theorem'' \cite{Feinberg1972}. This is the~history.

In our work, we develop the Michaelis-Menten-Stueckelberg approach
to general kinetics. This is a combination of the QE (fast
equilibria) and the QSS (small amounts) approaches to the real or
hypothetical intermediate states. These intermediate states
(compounds) are included in all elementary reactions
(\ref{stoichiometricequationcompaund}) as it is illustrated in
Figure~\ref{n-tail}. Because of the small amount, the free energy
for these compounds $B_i$ is perfect (\ref{FreeEn1}), the kinetics
of compounds is the first order Markov kinetics and satisfies the
master equation.

After that, we use the combination of QE and QSS approximations and
exclude the concentrations of compounds. For the general kinetics
the main result of this approach is the general kinetic law
(\ref{GeneralReactionRate}). Earlier, we just postulated this law
because of its convenient and natural form \cite{G1,BykGOrYab1982},
now we have the physical framework where this law can be proved.

We do not assume anything about reaction rates of the main reactions
(\ref{stoichiometricequation}). We use only thermodynamic
equilibrium, the hypothesis about fast equilibrium with compounds
and the smallness of concentration of compounds. This smallness
implies the perfect entropy and the first order kinetics for
compounds. After that, we get the reaction rate functions from the
qualitative assumptions about compounds and the equilibrium
thermodynamic data.

For example, if we relax the assumption about fast equilibrium and
use just smallness of compound concentrations (the Briggs-Haldane
QSS approach \cite{BriggsHaldane1925,Aris1965,Segel89}) then we
immediately need the formulas for reaction rates of compound
production. Equilibrium data become insufficient. If we relax the
assumption about smallness of concentrations then we lose the
perfect entropy and the first order Markov kinetics. So, only the
combination of QE and QSS gives the desired result.

For the kinetics of rarefied gases the mass action law for elastic
collisions (the Boltzmann equation) or for inelastic processes like
chemical reactions follows from the ``molecular chaos'' hypothesis
and the low density limits. The Michaelis-Menten-Stueckelberg
approach substitutes low density of all components by {\em low
density of the elementary events} (or of the correspondent
compounds) together with the QE~assumption.

The general kinetic law has a simple form: For an elementary
reaction $$\sum_i \alpha_i A_i \to \sum_i \beta_i A_i$$ the
reaction rate is $r=\varphi \Omega$, where $\Omega>0$ is the
Boltzmann factor, $\Omega=\exp\left(\sum_i \alpha_i
\check{\mu}_i\right)$, $\check{\mu}_i=-\partial S/ \partial
N_i$ is the chemical potential $\mu$ divided by $RT$, and
$\varphi \geq 0$ is the kinetic factor. Kinetic factors for
different reactions should satisfy some conditions. Two of them
are connected to the basic physics:
\begin{itemize}
\item{The detailed balance: The kinetic factors for
    mutually reverse reaction should coincide,
    $\varphi^+=\varphi^-$. This identity is proven for
    systems with microreversibility
    (Section~\ref{DetMicro}).}
\item{The complex balance: The sum of the kinetic factors
    for all elementary reactions of the form $\sum_i
    \alpha_i A_i \to \ldots$ is equal to the sum of the
    kinetic factors for all elementary reactions of the
    form $\ldots \to \sum_i \alpha_i A_i $
    (\ref{complexbalanceGENKIN}). This identity is proven
    for all systems under the
    Michaelis-Menten-Stueckelberg assumptions about
    existence of intermediate compounds which are in fast
    equilibria with other components and are present in
    small amounts.}
\end{itemize}

For the general kinetic law we studied several sufficient conditions
of accordance between thermodynamics and kinetics: Detailed balance,
complex balance and $G$-inequality.

In the practice of modeling, a kinetic model may, initially, do not
respect thermodynamic conditions. For these cases, we solved the
problem of whether it is possible to add a linear function to
entropy in order to provide agreement with the given kinetic model
and deformed thermodynamics. The answer is constructive
(Section~\ref{Sec:EntropyDeform}) and allows us to prove the general
algebraic conditions for the detailed and complex balance.

Finally, we have to mention that Michaelis, Menten and Stueckelberg
did not prove their ``big theorem''. Michaelis and Menten did not
recognize that their beautiful result of mass action law produced
from the equilibrium relations between substrates and compounds, the
assumption about smallness of compound concentrations and the
natural hypothesis about linearity of compound kinetics is a general
theorem. Stueckelberg had much more and fully recognized that his
approach decouples the Boltzmann $H$-theorem and the
microreversibility (detailed balance). This is important because for
every professional in theoretical physics it is obvious that the
microreversibility cannot be important necessary condition for the
$H$-theorem. Entropy production should be positive without any
relation to detailed balance (the proof of the $H$-theorem for
systems with detailed balance is much simpler but it does not
matter: Just the Markov microkinetics is sufficient for it).
Nevertheless, Stueckelberg did not produce the generalized mass
action law and did not analyze the general kinetic equation. Later,
Horn, Jackson and Feinberg approached the complex balance conditions
again and studied the generalized mass action law but had no
significant interest in the microscopic assumptions behind these
properties. Therefore, this paper is the first publication of the
Michaelis-Menten-Stueckelberg theorem.

\section*{Appendix}
\section*{1. Quasiequilibrium Approximation \label{App1}}

\subsection*{1.1. Quasiequilibrium Manifold}

Let us consider a system in a domain $U$ of a real vector space $E$
given by differential equations
\begin{equation}\label{sys1A}
\frac{\D x}{\D t}=F(x)
\end{equation}
We assume that for any $x_0 \in U$, solution $x(t;x_0)$ to the
initial problem $x(0)=x_0$ for (\ref{sys1A}) exists for all $t>0$
and belongs to $U$. Shifts in time, $x_0 \mapsto x(t;x_0)$ ($t>0$),
form a semigroup in $U$.

We do not specify the space $E$ here. In general, it may be any
Banach or even more general space. For nonlinear operators we will
use the Fr\'eshet differentials: For an operator $\Psi(x)$ the
differential at point $x$ is a linear operator $(D\Psi)_x$:
$$(D\Psi)_x (y)= \left. \frac{\D \Psi(x+\alpha y)}{\D y}
\right|_{\alpha=0} $$ We use also notation $(D_x\Psi)$ when it is
necessary to stress the choice of independent variable. The choice
of variables is not obvious.

The QE approximation for (\ref{sys1A}) uses two basic entities:
entropy and slow variables.

{\em Entropy} $S$ is a concave Lyapunov function  with
non-degenerated Hessian for (\ref{sys1}) which increases in~time:
\begin{equation}\label{SecondLawA}
\frac{\D S}{\D t}\geq 0  \,
\end{equation}
In this approach, the increase of the entropy in time is exploited
(the Second Law in the form (\ref{SecondLaw})).

Formally, any Lyapunov function may be used. Nevertheless, most of
famous entropies, like the relative Boltzmann-Gibbs-Shannon entropy,
the R\'enyi entropy, the Burg entropy, the Cressie-Read and the
Tsallis entropies could be defined as {\em universal} Lyapunov
functions for Markov chains which satisfy some natural additivity
conditions \cite{GorbanGorbanJudge2010}.

``Universal'' means that they do not depend on kinetic coefficients
directly but only on the equilibrium point. The ``natural additivity
conditions'' require that these entropies can be represented by sums
(or integrals) over states maybe after some monotonic transformation
of the entropy scale, and, at the same time, are additive with
respect to the joining of statistically independent systems (maybe,
after some monotonic rescaling as well).

{\em Slow variables} $M$ are defined as some differentiable
functions of variables $x$: $M=m(x)$. We use notation $E_M$ for the
space of slow variables, $M \in E_M$. Selection of the slow
variables implies a hypothesis about separation of fast and slow
motion. In its strongest form it consists of two assumptions: The
slaving assumption and the assumption of small fast-slow projection.

\vspace{12pt} \noindent{\bf The slaving assumption.} For any
admissible initial state $x_0 \in U$ after some relatively small
time $\tau$ ({\em initial layer}), solution $x(t;x_0)$ becomes a
function of $M$ (up to a given accuracy $\epsilon$) and can be
represented in a slaving form:
\begin{equation}\label{SlavingA}
x(t)=x^*_{M(t)}+\delta(t) \;\; {\rm for} \;\; t>\tau\, , \;\; {\rm
where} \; \; M(t)=m(x(t)),\; \|\delta(t)\|<\epsilon \,
\end{equation}
This means that everything is a function of slow variables, after
some initial time and up to a given~accuracy.

The smallness of $\tau$ is essential. If there is no restriction on
$\tau$ then every globally stable system will satisfy this
assumption because after some time it will arrive into a small
vicinity of equilibrium.

The second assumption requires that the slow variables (almost) do
not change during the fast motion: During the initial layer $\tau$,
the state $x$ can change significantly because of fast motion, but
the change in $M=m(x)$ during $\tau$ are small with $\tau$: {\bf The
assumption of small fast-slow projection}.

The QE approximation defines the functions $x^*_{M}$ as solutions to
the following {\bf MaxEnt} optimization~problem:
\begin{equation}\label{MaxEntA}
S(x) \to \max \;\; \mbox{subject to} \;\; m(x)=M \,
\end{equation}
The reasoning behind this approximation is simple: During the fast
initial layer motion, entropy increases and $M$ almost does not
change. Therefore, it is natural to assume (and even to prove using
smallness of $\tau$ and $\epsilon$ if the entropy gradient in fast
directions is separated from zero) that $x^*_{M}$ in
(\ref{SlavingA}) is close to solution to the MaxEnt optimization
problem (\ref{MaxEntA}). Further,  $x^*_{M}$ denotes a solution to
the MaxEnt problem.

Some additional conditions on $m$ and $S$ are needed for the
regularity of the dependence $x^*_{M}$ on $M$. It is more convenient
to discuss these conditions separately for more specific systems. In
general settings, let us just assume that for given $S$ and $m$ the
dependencies $m(x)$ and $x^*_{M}$ are differentiable. For their
differentials we use the notations $(Dm)_{x}$ and $(Dx^*_M)_M$. The
differentials are linear operators: $(Dm)_{x}: E \to E_M$ and
$(Dx^*_M)_M: E_M \to E$.

A solution to (\ref{MaxEntA}), $x^*_{M}$, is the {\em QE state}, the
corresponding value of the entropy
\begin{equation}\label{QEentropyA}
S^*(M)=S(x^*_{M})
\end{equation}
is the {\em QE entropy} and the equation
\begin{equation}\label{QEequationA}
\frac{\D M}{\D t}= (Dm)_{x^*_M} (F(x^*_M))
\end{equation}
represents the {\em QE dynamics}.

\vspace{12pt}\noindent{\it{\bf Remark.} The strong form of the
slaving assumption, ``everything becomes a function of the slow
variables'', is too strong for practical needs. In practice, we need
just to have a ``good'' dependence on $M$ for the time derivative
$\D M/ \D t$. Moreover, the short-time fluctuations of $\D M/ \D t$
do not affect the dependence $M(t)$ too much, and only the average
values $$\langle \dot{M}\rangle _{\theta}(t)= \frac{
1}{\theta}\int_t^{t+\theta} \frac{\D M}{\D t}
$$
for sufficiently small time scale $\theta$ are
important.}\vspace{2mm}

\subsection*{1.2. Preservation of Entropy  Production
\label{PreservEntrProdA}}

\noindent{\bf Theorem about preservation of entropy production.}
{\it Let us calculate  $\D S^*(M)/ \D t$ at point $M$ according to
the QE dynamics (\ref{QEequationA}) and find $\D S(x)/ \D t$ at
point $x=x^*_M$ due to the initial system~(\ref{sys1}). The results
always coincide:
\begin{equation}\label{PreservationEntProdA}
\frac{\D S^*(M)}{\D t}=\frac{\D S(x)}{\D t} \,
\end{equation}}

The left hand side in (\ref{PreservationEntProdA})  is computed due
to the QE approximation (\ref{QEequationA}) and the right hand side
corresponds to the initial system (\ref{sys1A}). Here, this theorem
is formulated in more general setting than in Section \ref{General}:
The slow variables $M=m(x)$ may be nonlinear functions of $x$. For
more details about QE approximation with nonlinear dependencies
$M=m(x)$ we refer to papers
\cite{GorKarQE2006,GorKarProjector2004,GorKarPRE1996}. The general
theorem about preservation of entropy production and thermodynamic
projector is presented in \cite{GorKarProjector2004}.

\begin{proof}To prove this identity let us mention that
\begin{equation}
\frac{\D S^*(M)}{\D t} = (DS^*)_M \left(\frac{\D M}{\D t}\right)
=(DS^*)_M \circ (Dm)_{x^*_M} (F(x^*_M))\,
\end{equation}
where $\circ$ stands for superposition. On the other hand, just from
the definitions of the differential and of the time derivative of a
function due to a system of differential equations, we get
\begin{equation}
\frac{\D S(x)}{\D t}=(D S)_x (F(x)) \,
\end{equation}
To finalize the proof, we need an identity
\begin{equation}\label{QEEntropyIdentityA}
(DS^*)_M \circ (Dm)_{x^*_M}=(D S)_{x^*_M} \,
\end{equation}

Let us use the Lagrange multipliers representation of the MaxEnt
problem:
\begin{equation}\label{LagrangeFormA}
(DS)_x= \Lambda_M \circ (Dm)_x \, , \;\; m(x)=M \,
\end{equation}
This system of two equations has two unknowns: The vector of state
$x$ and the linear functional $\Lambda_M$ on the space of slow
variables (the Lagrange multiplier), which depends on $M$ as on a
parameter.

By differentiation of the second equation $m(x)=M$, we get an
identity
\begin{equation}\label{normalizationDiffA}
(Dm)_{x^*_M} \circ(Dx^*_M)_M={\rm id} _{E_M} \,
\end{equation}
where ${\rm id}$ is the unit operator.

Lagrange multiplier $\Lambda_M$ is the differential of the QE
entropy:
\begin{equation}\label{LagrangeEntropyA}
(DS^*)_M=\Lambda_M \,
\end{equation}
Indeed, due to the chain rule, $(DS^*)_M=(DS)_{x^*_M}\circ
(Dx^*_M)_M$, due to (\ref{LagrangeFormA}), $(DS)_x= \Lambda_M
(Dm)_x$ \linebreak and, finally
\begin{equation*}
\begin{split}
(DS^*)_M=&(DS)_{x^*_M} \circ(Dx^*_M)_M=\Lambda_M \circ(Dm)_x
\circ(Dx^*_M)_M \\ =&\Lambda_M\circ {\rm id} _{E_M}=\Lambda_M \,
\end{split}
\end{equation*}

Now we can prove the identity (\ref{QEEntropyIdentityA}):
$$(DS^*)_M \circ (Dm)_{x^*_M}=\Lambda_M \circ(Dm)_{x^*_M}=(D S)_{x^*_M}$$
(here we use the Lagrange multiplier form (\ref{LagrangeFormA})
again).
\end{proof}

The preservation of the entropy production leads to the {\em
preservation of the type of dynamics}: If for the initial system
(\ref{sys1A}) entropy production is non-negative, $\D S/\D t \geq
0$, then for the QE approximation~(\ref{QEequationA}) the production
of the QE entropy is also non-negative, $\D S^*/\D t \geq 0$.

In addition, if for the initial system $({\D S}/{\D t})_x = 0$ if
and only if $F(x)=0$ then the same property holds in the QE
approximation.

\section*{2. First Order Kinetics and Markov Chains \label{App2}}

First-order kinetics form the simplest and well-studied class of
kinetic systems. It includes the continuous-time Markov chains (the
master equation \cite{VanKampen1981}), kinetics of monomolecular and
pseudomonomolecular reactions \cite{LumpWei2}, and has many other
applications.

We consider a general network of linear reactions. This network is
represented as a directed graph (digraph)
(\cite{Yablonskii1991,Temkin1996}): Vertices correspond to
components $B_j$ ($1 \leq j \leq q$), edges correspond to reactions
$B_j \to B_l$. For each vertex $B_j$ a positive real variable
$\varsigma_j$ (concentration) is defined. For each reaction $B_j \to
B_l$ a rate constant $\kappa_{lj} >0$ is given. To follow the
standard notation of the matrix multiplication, the order of indexes
in $\kappa_{ji}$ is always inverse with respect to reaction: It is
$\kappa_{j\leftarrow i}$, where the arrow shows the direction of the
reaction. The kinetic equations for concentrations $\varsigma_j$
have the form
\begin{equation}\label{MasterEqA}
\frac{\D \varsigma_j}{\D t}=\sum_{l, \, l\neq j} \left(\kappa_{jl}
\varsigma_l-\kappa_{lj}\varsigma_j\right)
\end{equation}

The linear conservation law (for the Markov chains this is the
conservation of the total probability) is:

\begin{equation}\label{MasterEqBalanceA}
\sum_j \frac{\D \varsigma_j}{\D t}=0 \;\; \mbox{{\em i.e.},} \;\;
\sum_{j,l, \, l\neq j} \left(\kappa_{jl}
\varsigma_l-\kappa_{lj}\varsigma_j\right)=0\, .
\end{equation}

Let a positive vector $\varsigma^*$ ($\varsigma^*_j >0$) be an
equilibrium for the system (\ref{MasterEqA}): For every $j=1, \ldots
, q$
\begin{equation}\label{MarkovEqEquilibriumCondA}
\sum_{l, \, l\neq j} \kappa_{jl} \varsigma_l^*=\sum_{l, \, l\neq j}
\kappa_{lj}\varsigma_j^*
\end{equation}

An equivalent form of (\ref{MasterEqA}) is convenient. Let us use
the equilibrium condition (\ref{MarkovEqEquilibriumCondA}) and write
$$\sum_{l, \, l\neq j}\kappa_{lj}\varsigma_j=\left(\sum_{l, \, l\neq
j}\kappa_{lj}\varsigma_j^*\right)\frac{\varsigma_j}{\varsigma_j^*}=
\sum_{l, \, l\neq j} \kappa_{jl}
\varsigma_l^*\frac{\varsigma_j}{\varsigma_j^*} $$ Therefore, under
condition (\ref{MarkovEqEquilibriumCondA}) the master equation
(\ref{MasterEqA}) has the equivalent form:
\begin{equation}\label{EquivalentMarkovA}
\frac{\D \varsigma_j}{\D t}=\sum_{l, \, l\neq j} \kappa_{jl}
\varsigma_l^*\left(\frac{\varsigma_l}{\varsigma_l^*}-\frac{\varsigma_j}{\varsigma_j^*}\right)
\end{equation}

The following theorem \cite{Morimoto1963} describes the large class
of the Lyapunov functions for the first order kinetics. Let $h(x)$
be a smooth convex function on the positive real axis. A
Csisz\'ar--Morimoto function $H_h(\varsigma)$ is (see the review
\cite{GorbanGorbanJudge2010}):
 $$H_h(\varsigma)=\sum_l \varsigma_l^*
 h\left(\frac{\varsigma_l}{\varsigma_l^*}\right)$$

\vspace{2mm}\noindent{\bf The Morimoto $H$-theorem.} {\it The time
derivative of $H_h(\varsigma)$ due to (\ref{MasterEqA}) under
condition (\ref{MarkovEqEquilibriumCondA}) is~nonpositive:
\begin{equation}\label{ENtropyProdA}
\frac{\D H_h(\varsigma)}{\D t}= \sum_{l,j, \, j\neq l}
h'\left(\frac{\varsigma_j}{\varsigma_j^*}\right)
\kappa_{jl}\varsigma^*_l \left(\frac{\varsigma_l}{\varsigma _l^*}-
\frac{\varsigma_j}{\varsigma_j^*}\right) \leq 0
\end{equation}}

\vspace{2mm}
\begin{proof}Let us mention that for any $q$ numbers $h_i$, $\sum_{i,j, \, j\neq i}
\kappa_{ij}\varsigma^*_j(h_j-h_i)=0$. Indeed, for $h_i=p_i/p_i^*$
this is precisely the condition of conservation of the total
probability for equations (\ref{EquivalentMarkovA}). The extension
from a simplex of the $h_i$ values ($\sum_i p^*_i h_i=1$, $h_i\geq
0$) to the positive orthant $\mathbb{R}^q_+$ is trivial because of
uniformity of the identity. Finally, if a linear identity holds in a
positive orthant then it holds in the whole space $\mathbb{R}^q$.
Therefore,
$$\sum_{l,j, \, j\neq l} h'\left(\frac{\varsigma_j}{\varsigma_j^*}\right)
\kappa_{jl}\varsigma^*_l \left(\frac{\varsigma_l}{\varsigma _l^*}-
\frac{\varsigma_j}{\varsigma_j^*}\right) = \sum_{l,j, \, j\neq l}
\kappa_{jl}\varsigma^*_l\left[h\left(\frac{\varsigma_j}
{\varsigma_j^*}\right)-
h\left(\frac{\varsigma_l}{\varsigma_l^*}\right) +
h'\left(\frac{\varsigma_j}{\varsigma_j^*}\right)\left(\frac{\varsigma_l}{\varsigma
_l^*}- \frac{\varsigma_j}{\varsigma_j^*}\right)\right] \leq 0
$$
The last inequality  holds because of the convexity of $h(x)$:
$h'(x)(y-x)\leq h(y)-h(x)$ \linebreak (Jensen's
inequality).\end{proof}

For example, for the convex function $h(x)=x(\ln x-1)$ the
Csisz\'ar--Morimoto function is:
\begin{equation}\label{FreEnA}
H_h(\varsigma)=\sum_l \varsigma_l
\left(\ln\left(\frac{\varsigma_l}{\varsigma^*_l}\right)-1\right)
\end{equation}
This expression coincides with the perfect component of the free
energy (\ref{FreeEn2}) (to be more precise, \linebreak  $f=RT
H_h(\varsigma)$).

Each positive equilibrium $\varsigma^*$ belongs to the ray of
positive equilibria $ \lambda \varsigma^*$ ($\lambda>0$). We can
select a $l_1$-normalized direction vector and write for an
equilibrium $\varsigma^{\rm eq}$ from this ray:
$$\varsigma^{\rm eq}= \beta \frac{\varsigma^*}{\sum_j
\varsigma^*_j}\, ,$$ where $\beta=\sum_j \varsigma^{\rm eq}_j$.

The kinetic equations (\ref{MasterEqA}) allow one and only one ray
of positive equilibria if and only if the digraph of reactions is
strongly connected:  It is possible to reach any vertex starting
from any other vertex by traversing edges in the directions in which
they point. Such continuous-time Markov chains are called ergodic
chains \cite{MeynMarkCh2009}.

Let us assume that the system is {\em weakly reversible}: For any
two vertices $B_i$ and $B_j$, the existence of an oriented path from
$B_i$ to $B_j$ implies the existence of an oriented path from $B_j$
to $B_i$. Under this assumption the graph of reactions is a unit of
strongly connected subgraphs without connections between them. Let
the graph of reactions $B_j \to B_l$ have $d$ strongly connected
components $C_s$ and let $V_s$ be the set of indexes of those $B_j$
which belong to $C_s$: $B_j \in C_s$ if and only if $j \in V_s$. For
each $s=1, \ldots, d$ there exists a conservation~law
\begin{equation}\label{LinearConservationA}
\beta_s(\varsigma)=\sum_{j\in V_s}\varsigma_j=const
\end{equation}
For any set of positive values of $\beta_s>0$ ($s=1, \ldots , d$)
there exists a unique equilibrium of (\ref{MasterEqA}),
$\varsigma^{\rm eq}$, which is positive ($\varsigma^{\rm eq}_j>0$).
This equilibrium can be expressed through any positive equilibrium
$\varsigma^*$:
\begin{equation}\label{compoundEqA}
\varsigma^{\rm eq}_j=\beta_s\frac{\varsigma_j^*(c,T)}{\sum_{l\in V_s
}\varsigma_j^*(c,T)}
\end{equation}

For positive $\beta_s$, logarithms of $\varsigma^{\rm eq}$ form a
$d$-dimensional linear manifold in $R^q$:
\begin{equation}\label{complexMarkovEq1A}
\ln \varsigma^{\rm eq}_j=\ln \beta_s + \ln
\left(\frac{\varsigma_j^*(c,T)}{\sum_{l\in V_s
}\varsigma_l^*(c,T)}\right)
\end{equation}
The natural coordinates on this manifold are $\ln \beta_s$.

\vspace{12pt}\noindent{\it {\bf Remark}. In the construction of the
free energy (\ref{FreEnA}) any positive equilibrium state
$\varsigma^{\rm eq}$ can be used. The correspondent functions
differs in an additive constant. Let us calculate the difference
between two ``free energies'':
\begin{equation}
\begin{split}
&\sum_{j=1}^q \varsigma_j
\left(\ln\left(\frac{\varsigma_j}{\varsigma_j^*}\right)-1\right)-\sum_{j=1}^q
\varsigma_j \left(\ln\left(\frac{\varsigma_j}{\varsigma_j^{\rm
eq}}\right)-1\right)\\&=\sum_{j=1}^q\varsigma_j
\ln\left(\frac{\varsigma_j^{\rm
eq}}{\varsigma_j^*}\right)=\sum_{s=1}^d
\beta_s(\varsigma)\ln\left(\frac{\beta_s(\varsigma^{\rm
eq})}{\beta_s(\varsigma^*)}\right)
\end{split}
\end{equation}
The result is constant in time for the solutions of the master
equation (\ref{MasterEqA}), hence, these functions are equivalent:
They both are the Lyapunov functions for (\ref{MasterEqA}) and have
the same conditional minimizers for given values of $\beta_s>0$
($s=1, \ldots , d$).}\vspace{2mm}

\end{document}